\documentclass[sigconf,nonacm,balance=false]{acmart}
\usepackage{amsmath,booktabs,multirow,float}
\setcopyright{none}
\renewcommand\footnotetextcopyrightpermission[1]{}
\newcommand{\Fone}{\ensuremath{F_1}}

\makeatletter
\if@ACM@nonacm
  \newcommand{\preprintpagestyle}{\fancyhf{}\fancyfoot[C]{\footnotesize\thepage}\renewcommand{\headrulewidth}{0pt}\renewcommand{\footrulewidth}{0pt}}
  \AtBeginDocument{\fancypagestyle{standardpagestyle}{\preprintpagestyle}\fancypagestyle{firstpagestyle}{\preprintpagestyle}\pagestyle{standardpagestyle}}
\fi
\makeatother
\begin{document}
\title[MuQ-LC: Long-Context Guzheng Playing Technique Detection]{MuQ-LC: Long-Context Music Representation Adaptation for Guzheng Playing Technique Detection}
\author{Youwei Feng}
\authornote{These authors contributed equally to this work and share first authorship.}
\email{fengyw01@zenmen.com}
\affiliation{%
  \institution{Shanghai Lianshang Network Technology Group Co., Ltd.}
  \city{Shanghai}
  \country{China}
}

\author{Haoxuan Liu}
\authornotemark[1]
\authornote{Corresponding author.}
\email{liuhx02@zenmen.com}
\affiliation{%
  \institution{Shanghai Lianshang Network Technology Group Co., Ltd.}
  \city{Shanghai}
  \country{China}
}

\author{Yinquan Chen}
\email{chenyq01@zenmen.com}
\affiliation{%
  \institution{Shanghai Lianshang Network Technology Group Co., Ltd.}
  \city{Shanghai}
  \country{China}
}

\renewcommand{\shortauthors}{Feng et al.}
\begin{abstract}
Instrument playing technique detection requires both recognition of fine-grained acoustic activity and localization of individual performance gestures. In guzheng music, techniques exhibit heterogeneous temporal characteristics and may overlap or extend beyond note boundaries. Although pretrained music encoders support technique recognition, the contribution of downstream context length and its interaction with event localization remain important questions. This study introduces MuQ-LC, a framework that adapts a pretrained MuQ encoder for continuous multi-label technique recognition, extends its contextual input using concatenated short-segment representations, and aggregates overlapping predictions in logit space. A separately optimized, class-agnostic technique-onset head subsequently localizes events while retaining the established activity predictions. On the Guzheng Tech99 official split, increasing the encoder context from 3 to 18 seconds and introducing overlapping inference improves test frame macro-\Fone{} from 82.89\% to 88.11\%, with consistent improvements across three continuation-training seeds. Decoupled onset learning raises event macro-\Fone{} from 79.18\% to 81.06\% relative to a note-onset baseline while preserving the activity predictions. The resulting system scores 3.61 percentage points above the strongest published MERTech frame macro-\Fone{} and 4.96 points above its full-system event macro-\Fone{}. Layer-wise adaptation experiments, an eight-length context sweep, and candidate-level diagnostics characterize the respective contributions of representation adaptation, temporal context, and onset learning.
\end{abstract}
\maketitle

\section{Introduction}
\label{sec:intro}
Instrument playing techniques (IPTs) encode how a musical sound is produced and shaped, complementing conventional descriptors such as pitch, onset, and duration. In the guzheng, a Chinese plucked-string instrument, techniques include plucking, pitch modulation through string manipulation, and repeated excitation that produces a sustained texture. Such gestures provide useful descriptors for performance analysis, transcription, and retrieval of musical examples. Their identification is particularly challenging in continuous recordings: techniques may overlap, a single gesture may extend across several notes, and acoustically similar local observations may correspond to different actions. Related studies of guzheng, guqin, and guitar performance have established the importance of temporal and musical context for interpreting these gestures \cite{li2022guzheng,li2023tech99,huang2020guqin,su2019tent}.

The development of guzheng IPT detection has progressed from onset-assisted recognition to continuous, frame-level multi-label prediction. The GZ IsoTech study combined technique classification with note-onset information \cite{li2022guzheng}; the subsequent Tech99 study introduced multi-scale processing and self-attention to address the heterogeneous temporal characteristics of techniques \cite{li2023tech99}. MERTech further demonstrated the utility of pretrained music representations and joint supervision for pitch, technique activity, and technique onsets \cite{li2024mertech}. Recent violin research likewise exploits multi-scale temporal experts and pretrained music features \cite{wang2026viotech}. These approaches establish temporal modeling and onset supervision as central components of technique detection.

Three related design questions motivate the present study. First, a representation that supports general music understanding does not necessarily provide the most appropriate readout for dense technique recognition. Layer-wise investigations and source-aware probing of Chinese and Western music reveal substantial task dependence in pretrained representations \cite{kanatas2026layer,zhou2025layerwise,bai2026dimension}. Second, the temporal receptive field of a local prediction head and the amount of input context available to a pretrained encoder are distinct design variables. Extending encoder context may provide information about surrounding musical gestures without changing the local prediction head. Third, recognizing activity and localizing event starts impose different requirements: a small temporal displacement may have little effect on frame-level overlap but invalidate an onset match. Onset-informed transcription and beat tracking illustrate the value of examining these objectives separately \cite{hawthorne2018onsets,foscarin2024beatthis}. MERTech's Tech99 ablation makes this distinction particularly clear: onset-conditioned post-processing increases event macro-\Fone{} from 57.0\% to 76.1\%, while frame macro-\Fone{} decreases from 84.5\% to 80.4\% \cite{li2024mertech}. This trade-off motivates an activity-preserving event detector that improves localization without replacing the dense activity output.

MuQ-LC addresses these questions through a staged approach. A frozen MuQ acoustic frontend \cite{zhu2025muq} produces short-segment representations, while the downstream Conformer encoder is adapted for multi-label technique activity. Adjacent representations are concatenated to extend the context available to the encoder, retaining the original local temporal readout. Overlapping inference combines estimates of each frame under multiple contexts before converting the aggregated logits into activity probabilities. Subsequently, a shared technique-onset head is optimized on the adapted contextual features with the activity recognizer fixed. The onset predictions are associated with the unchanged activity probabilities to produce labeled technique events.

The contributions are organized around three research questions:
\begin{itemize}
\item \textbf{Representation adaptation.} Layer probing, temporal readout, and encoder adaptation experiments characterize the requirements of continuous fine-grained technique recognition and motivate the task-specific activity model.
\item \textbf{Long-context recognition.} MuQ-LC combines extended downstream encoder context with overlapping logit aggregation. An eight-length sweep and paired continuation experiments separate the context and aggregation gains, yielding a 5.22 percentage-point improvement in test frame macro-\Fone{} over the short-context baseline.
\item \textbf{Decoupled event localization.} A separately trained shared technique-onset head exploits the adapted representation without modifying the activity predictions. Candidate diagnostics and comparisons with joint adaptation characterize event-level improvements; the decoupled configuration improves event macro-\Fone{} by 1.89 percentage points over note-onset anchoring.
\end{itemize}

\section{Related Work}
\label{sec:related}
\subsection{Instrument Playing Technique Detection}
Early technique recognition methods exploit instrument-specific acoustic behavior, including pitch contours for electric guitar techniques \cite{chen2015guitar} and fully convolutional processing of variable-length bowed-string recordings \cite{wang2019erhu}. Technique-aware transcription provides a complementary perspective: TENT incorporates guitar technique information into note tracking \cite{su2019tent}, while recent embedded acoustic-guitar recognition addresses continuous performance in an interactive setting \cite{stefani2025guitar}. These studies illustrate the relationship between musical gestures and note-level representations, rather than treating technique labels solely as recording-level attributes.

For guzheng, note-onset-assisted detection uses note boundaries to integrate technique predictions \cite{li2022guzheng}. Tech99 formulates the problem as frame-level multi-label recognition and combines multi-scale convolution with self-attention \cite{li2023tech99}. Its analysis of technique-dependent duration motivates temporal modeling at multiple scales. MERTech processes 5-s waveforms using a learned mixture of MERT layers, jointly predicts pitch and technique activity with a pitch--technique tensor, and incorporates a class-agnostic IPT-onset branch into attention-based refinement and event decoding \cite{li2024mertech}. More recently, VioPTT incorporates technique labels into violin transcription \cite{wang2026vioptt}, and VioTech combines frozen MERT representations, note-related cues, and multi-scale temporal experts for violin activity recognition \cite{wang2026viotech}. MuQ-LC instead extends the adapted encoder context to 18 seconds while retaining the local temporal readout, and separates activity learning from subsequent onset optimization and event decoding.

\subsection{Music Foundation Models and Task Adaptation}
Self-supervised and supervised audio representation learning has established a foundation for transferring musical features to downstream tasks \cite{mccallum2022representations,yuan2023marble}. MERT uses acoustic and musical targets during large-scale self-supervised training \cite{li2024mert}; MusicFM provides transferable representations for music informatics \cite{won2024musicfm}; MuQ learns masked music representations using mel residual vector quantization \cite{zhu2025muq}. These models provide a useful starting point for annotation-limited, instrument-specific tasks.

The usefulness of pretrained features depends on the musical domain, representation layer, and adaptation strategy. Studies of world music corpora \cite{papaioannou2025universal} and continual pretraining \cite{kanatas2025culturemert} investigate cross-cultural transfer, while work on Iranian classical music compares alternative task adaptation procedures \cite{mohseni2026iranian}. Layer-wise analyses characterize differences in the information accessible from intermediate music representations \cite{kanatas2026layer,zhou2025layerwise}. In particular, dimension-wise probing of Chinese and Western music reports strong MuQ performance for source-aware isolated guzheng technique classification \cite{bai2026dimension}, providing an empirical foundation for its application to continuous technique detection. Building on these findings, MuQ-LC systematically examines layer selection and temporal readout, and adapts the contextual encoder through dense multi-label supervision to support frame-level recognition in continuous performances.

\subsection{Temporal Modeling and Event Localization}
Temporal structure is central to MIR tasks that require localized decisions. Joint analysis of mode and technique in guqin performance demonstrates the relevance of broader musical information to local gesture recognition \cite{huang2020guqin}. Beat Transformer models long-range temporal relations using dilated self-attention \cite{zhao2022beattransformer}, while Beat This! investigates accurate beat tracking without conventional DBN post-processing \cite{foscarin2024beatthis}. Hierarchical frequency--time modeling and sparse-attention approaches to piano transcription further illustrate the importance of contextual representations and input length \cite{toyama2023hft,wei2025sparse}. MuQ-LC extends this line of research to continuous playing technique detection by adapting a pretrained music encoder to longer input sequences and integrating predictions across overlapping temporal contexts.

Activity and onset supervision are established components of music transcription. Onsets and Frames combines note-onset and frame predictions \cite{hawthorne2018onsets}, whereas high-resolution piano transcription explicitly regresses onset and offset times \cite{kong2021highresolution}. Event-based alternatives directly score intervals or decode structured note events \cite{yan2021semicrf,yan2024intervals}. Guzheng technique detection has advanced from note-onset anchoring to shared IPT-onset learning and onset-conditioned post-processing \cite{li2022guzheng,li2024mertech}. MuQ-LC separates both the optimization stages and the output paths: event decoding exploits the learned activity probabilities without reconstructing the activity sequence from decoded events.

\section{Method}
\label{sec:method}
\subsection{Problem Formulation}
Let $x$ denote a recording and $\mathcal C$ the seven technique classes. An annotated occurrence $e$ is represented by $(\tau_e,\omega_e,c_e)$, with onset $\tau_e$, offset $\omega_e$, and class $c_e$. All onset and offset times are measured in seconds. Let $T$ be the number of frames on the recording grid, indexed by $t=0,\ldots,T-1$, with center time $\operatorname{time}(t)=(t+1/2)/25$. The activity target at frame $t$ and class $c$ is
\begin{equation}
 y_{t,c}=\mathbf{1}\{\exists e:c_e=c,\ \tau_e\leq\operatorname{time}(t)<\omega_e\}.
 \label{eq:frame_target}
\end{equation}
The frame grid has a rate of 25~Hz, and the target matrix $Y=(y_{t,c})\in\{0,1\}^{T\times7}$ is multi-hot. The activity model predicts $P=(p_{t,c})\in[0,1]^{T\times7}$. The event task predicts a set $\widehat{\mathcal E}=\{(\hat\tau_n,\hat c_n)\}$ of labeled onsets; it does not require pitch or offset prediction. The two outputs represent complementary descriptions of a performance: the former specifies the duration of activity on the frame grid, and the latter identifies individual technique occurrences.

\begin{figure*}[t]
\centering
\includegraphics[width=\textwidth]{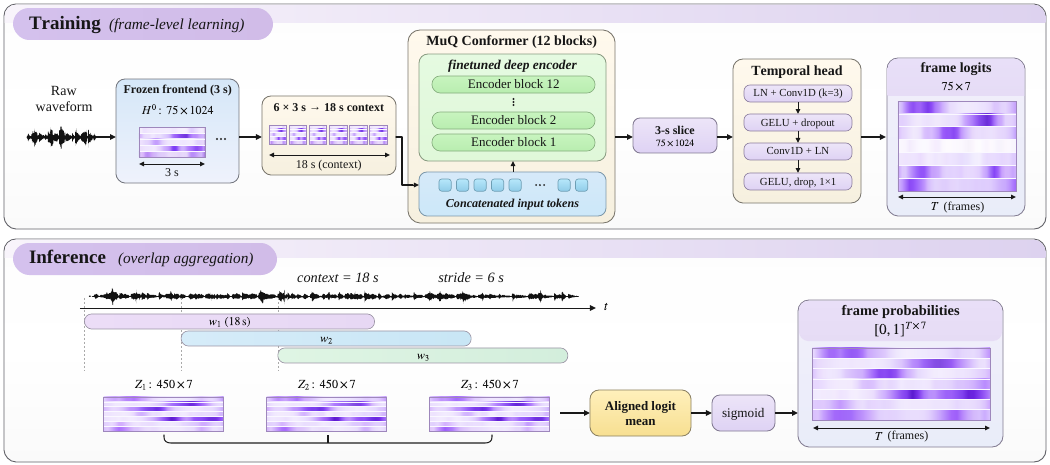}
\caption{Task-specific activity learning in MuQ-LC. The frozen MuQ frontend extracts 75-frame representations from individual 3-s audio segments. Concatenated features are processed by the adapted 12-block Conformer encoder; the contextual output is sliced into local sequences for the temporal head. The feature cells and waveform are schematic.}
\Description{A short-segment frontend, concatenated features, trainable Conformer encoder, local temporal convolutional head, and frame and clip supervision.}
\label{fig:architecture}
\end{figure*}

\subsection{Task-Specific Representation Adaptation}
The architecture in Fig.~\ref{fig:architecture} uses the \texttt{MuQ-large-msd-iter} checkpoint. MuQ's frozen acoustic frontend $F$ maps a non-overlapping 3-s audio segment $x_j$ to $H^0_j\in\mathbb R^{75\times1024}$ at 25~Hz. The 12 Conformer blocks and final encoder normalization are adapted under task supervision, while the frontend and positional module remain fixed. Conformer blocks combine self-attention and convolutional processing \cite{gulati2020conformer}, providing contextual representations without changing the frame grid.

The final-layer features are processed by a temporal convolutional head $g_\phi$. Its architecture consists of input layer normalization, a kernel-3 convolution from 1024 to 128 channels, GELU and dropout, a second kernel-3 convolution with 128 channels followed by layer normalization, GELU and dropout, and a kernel-1 projection to seven logits. The dropout rate is 0.15. The head operates on 75-frame slices even when the encoder input is longer, so that the local readout remains unchanged across context configurations.

Let $s_j$ be the start time of segment $j$, and $i=0,\ldots,74$ its local frame index. For a valid position ($v_{j,i}=1$), the recording index and target are
\begin{equation}
 t(j,i)=\operatorname{round}(25s_j)+i,\quad
 y_{j,i,c}=y_{t(j,i),c}.
 \label{eq:local_index}
\end{equation}
Invalid positions have $v_{j,i}=0$. For frame logits $z_{j,i,c}$ and positive-class weight $a_c$, define weighted binary cross-entropy as
\begin{equation}
 b(z,y;a)=-ay\log\sigma(z)-(1-y)\log(1-\sigma(z)).
 \label{eq:bce}
\end{equation}
An auxiliary clip logit $\xi_{j,c}$ is obtained from the masked maximum and mean of the frame logits:
\begin{equation}
 \xi_{j,c}=0.65\max_{i:v_{j,i}=1}z_{j,i,c}
 +0.35\frac{\sum_i v_{j,i}z_{j,i,c}}{\max(1,\sum_i v_{j,i})}.
 \label{eq:clip}
\end{equation}
Let $I_j=[s_j,s_j+D_j)$ be the valid audio interval of segment $j$, with duration $D_j$ in seconds. For class $c$, the union of annotated activity intervals is $\mathcal A_c=\bigcup_{e:c_e=c}[\tau_e,\omega_e)$. The clip label is
\begin{equation}
 \ell_{j,c}=\mathbf{1}\{|\mathcal A_c\cap I_j|\geq\max(0.05,0.01D_j)\},
 \label{eq:clip_target}
\end{equation}
where $|\cdot|$ denotes interval duration in seconds. The union counts overlapping same-class annotations only once. Frame- and clip-level positive weights are computed on training data as
\begin{equation}
 a_c^m=\operatorname{clip}\left(\sqrt{\frac{N^m-N_c^m}{\max(N_c^m,1)}},1,8\right),
 \quad m\in\{f,c\},
 \label{eq:activity_weights}
\end{equation}
where $N^m$ counts all valid frames ($m=f$) or all clips ($m=c$), and $N_c^m$ counts the corresponding positive labels for class $c$. The clipping operator restricts the weight to $[1,8]$. For a batch of $J$ segments, the objective is
\begin{align}
 \mathcal L_f &= \frac{\sum_{j,i,c} v_{j,i} b(z_{j,i,c},y_{j,i,c};a_c^f)}{7\max(1,\sum_{j,i}v_{j,i})},\\
 \mathcal L_c &= \frac{1}{7J}\sum_{j,c} b(\xi_{j,c},\ell_{j,c};a_c^c),\\
 \mathcal L_{\mathrm{act}} &= \mathcal L_f+0.25\mathcal L_c.
 \label{eq:actloss}
\end{align}
This objective combines time-localized activity supervision with clip-level evidence of technique presence. Layer probing and adaptation experiments motivating the final encoder configuration are presented in Sec.~\ref{sec:adaptresults}.

\begin{figure*}[t]
\centering
\includegraphics[width=\textwidth]{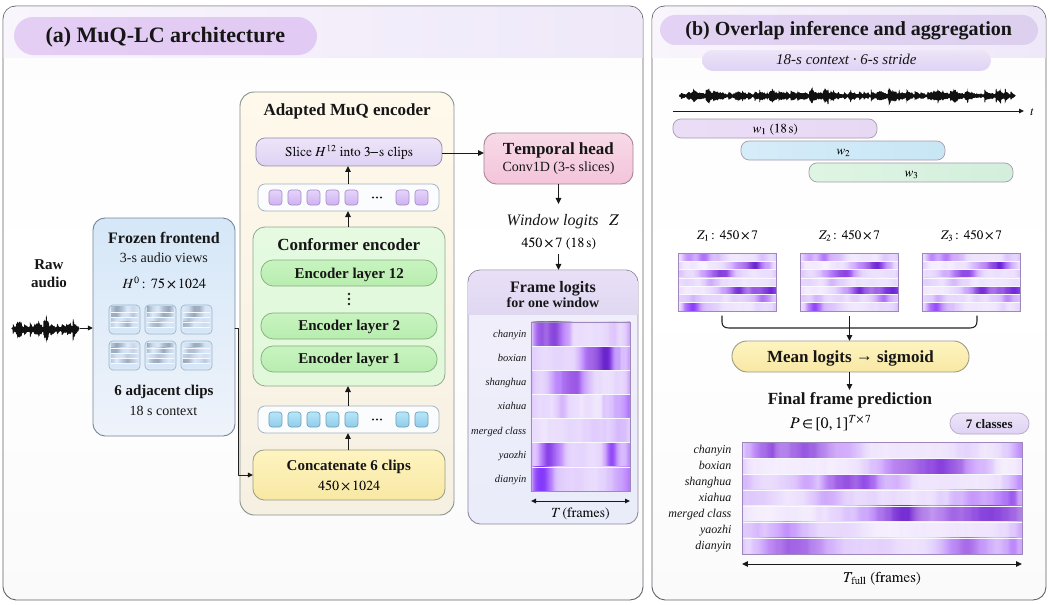}
\caption{Long-context modeling and overlapping inference. Consecutive 3-s features are grouped for 18-s encoder training. Inference reuses the trained model with a nominal 6-s stride, aligning each window's logits to the recording timeline. Logits from covering windows are averaged before sigmoid. The depicted windows and features are schematic.}
\Description{Non-overlapping short features form long training groups; overlapping inference windows provide aligned logits that are averaged before sigmoid.}
\label{fig:overlap}
\end{figure*}

\subsection{Long-Context Modeling}
Technique-dependent temporal cues may extend outside a short acoustic segment. The context of the adapted encoder is therefore extended by concatenating adjacent frontend representations within the same recording, as illustrated in Fig.~\ref{fig:overlap}. For a window $w$ consisting of $K$ consecutive segments,
\begin{align}
 H^0_j &= F(x_j)\in\mathbb R^{75\times1024},\\
 H^{(w)} &= E_\theta([H^0_{j_1};\ldots;H^0_{j_K}])\notag\\
 &\in\mathbb R^{75K\times1024},\\
 z^{(w)}_j &= g_\phi(\operatorname{slice}_j H^{(w)})\in\mathbb R^{75\times7}.
 \label{eq:context}
\end{align}
Here $E_\theta$ denotes the adapted contextual encoder. For segment $j_k$ at position $k=1,\ldots,K$ in the window,
\begin{equation}
 \operatorname{slice}_{j_k}H^{(w)}=H^{(w)}[75(k-1):75k,:],
 \label{eq:context_slice}
\end{equation}
where the row range uses zero-based, half-open indexing. The nominal 18-s context consists of six short representations, forming a $450\times1024$ input. This extends the encoder input, not the acoustic frontend's raw-waveform processing length. The encoder can relate frames across segment boundaries while the prediction head retains the original 3-s scope.

Activity training uses consecutive non-overlapping groups. Groups never cross recording boundaries, and discontinuities in segment timestamps terminate a chain. Short residual groups are processed using their actual segment counts rather than padding to six segments. Validity masks exclude invalid local frames from the activity readout and losses. The short-context baseline uses $K=1$ with the same initialization, head, and objective; the long-context configuration uses $K=6$. The context-length comparison thus evaluates a change in both training and non-overlapping inference context, with the local frontend and head architecture controlled.

\subsection{Overlapping Context Aggregation}
Non-overlapping inference assigns a frame a single contextual window. The position of that frame within the window determines which surrounding musical information is available to the encoder. Overlapping inference supplies additional contextual views without changing the learned parameters. MuQ-LC reuses the 18-s checkpoint with a nominal 6-s stride, corresponding to two short segments. A final shifted window covers the end of a chain when necessary, and chains shorter than 18~s use their available segments.

Let $\mathcal W_t$ denote the windows with valid predictions for recording-level frame $t$. For $w\in\mathcal W_t$, $z^{(w)}_{t,c}$ is the local logit mapped to $t$ by Eq.~\ref{eq:local_index}. For nonempty $\mathcal W_t$, the aggregated probability is
\begin{equation}
 p_{t,c}=\sigma\!\left(\frac{1}{|\mathcal W_t|}\sum_{w\in\mathcal W_t}z^{(w)}_{t,c}\right).
 \label{eq:overlap}
\end{equation}
The operation averages aligned logits rather than averaging already normalized probabilities. All contributions correspond to the same frame and class; invalid positions are excluded. Frames outside valid feature coverage are assigned zero probabilities when constructing the full recording output. This aggregation can combine estimates obtained from different surrounding contexts while retaining frame resolution. The comparison with non-overlapping inference uses the identical 18-s checkpoint and therefore isolates the effect of inference aggregation within the implemented system.

\begin{figure*}[t]
\centering
\includegraphics[width=\textwidth]{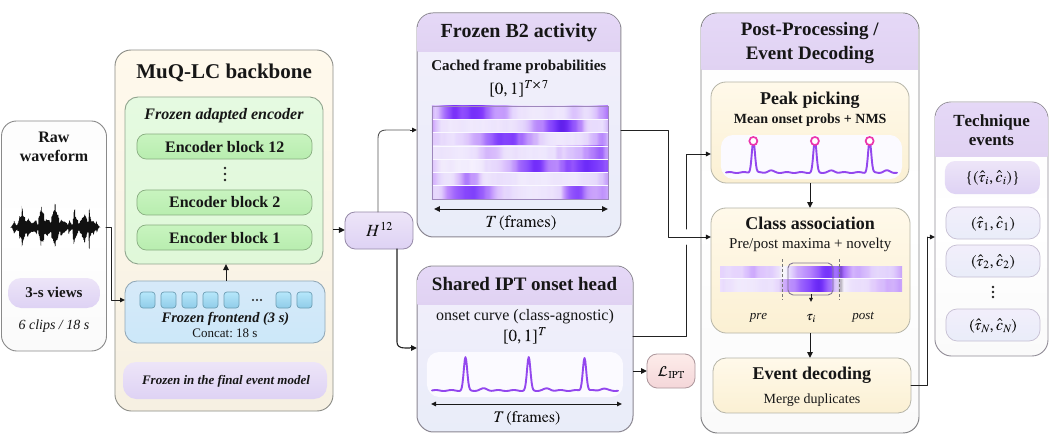}
\caption{Decoupled technique-onset localization. A trainable class-agnostic head predicts onset probabilities from frozen task-adapted contextual features. Overlapping onset probabilities are averaged and peak-picked; candidate times are associated with the unchanged activity cache using local pre/post evidence and a novelty criterion. The resulting output contains onset--class pairs. Feature grids and onset curves are illustrative.}
\Description{A frozen encoder feeds a trainable onset head; cached frame activity and peak-picked onset times feed an event association decoder.}
\label{fig:event}
\end{figure*}

\subsection{Decoupled Technique Event Localization}
The activity sequence describes where techniques are active but does not explicitly distinguish successive occurrences within an active region. Note attacks can provide candidate starts \cite{li2022guzheng}, yet a technique can begin during an existing note or continue across multiple attacks. Following the shared IPT-onset formulation in MERTech \cite{li2024mertech}, a single onset channel pools temporal supervision across technique classes. MuQ-LC optimizes this head after activity adaptation, with the activity model frozen, and maintains separate activity and event outputs (Fig.~\ref{fig:event}). This design retains dense activity estimates independently of onset-based event decisions.

\paragraph{Shared onset head.}
The final-layer representation from the adapted 18-s encoder feeds a head $a_\psi$ with an input projection (1024 to 128 channels), group normalization and GELU, four depthwise-separable temporal residual blocks with dilation rates 1, 2, 4, and 8, and a final 128-to-1 projection. A single logit per frame represents the likelihood of a technique start irrespective of class. Class identity is recovered during association with the activity sequence.

For reference onset $\tau_e$, the corresponding continuous recording-frame index is $\alpha_e=25\tau_e-1/2$. The recording-level soft target is a maximum over local Gaussian responses:
\begin{equation}
 q_t=\max_{e:t\in S_e}\exp\!\left[-\frac{(t-\alpha_e)^2}{2\cdot0.9^2}\right],
 \label{eq:target}
\end{equation}
where $S_e$ contains integer frames from $\lfloor \alpha_e-3.5\rfloor$ through $\lceil \alpha_e+3.5\rceil$, clipped to the recording, and an empty maximum is zero. The Gaussian standard deviation is 0.9 frames (36~ms). Max pooling retains a single target for coincident starts of different classes.

Onset training and inference use fixed 450-frame feature windows, with zero padding for short inputs and a validity mask excluding padded and uncovered positions from the losses and aggregation. For a window starting at recording-frame index $o$, local position $i=0,\ldots,449$ maps to $t=o+i$. Let $\nu$ enumerate the window-frame positions of the mini-batch, with $q_\nu$ the mapped target from Eq.~\ref{eq:target}, $\rho_\nu$ the corresponding rarity weight (Appendix~\ref{app:implementation}), $v_\nu$ the validity indicator, and $r_\nu=\sigma(u_\nu)$ the probability of onset logit $u_\nu$. The onset objective combines weighted BCE with soft Dice, aggregating over the entire mini-batch:
\begin{align}
 \mathcal L_{\mathrm{IPT}}&=\frac{\sum_\nu v_\nu[1+q_\nu(12\rho_\nu-1)]b(u_\nu,q_\nu;1)}{\max(1,\sum_\nu v_\nu)}\notag\\
 &\quad+0.1\left(1-\frac{2\sum_\nu v_\nu r_\nu q_\nu+1}{\sum_\nu v_\nu(r_\nu+q_\nu)+1}\right).
 \label{eq:iptloss}
\end{align}
The weight multiplies the BCE at the corresponding position, and the Dice term accounts for the sparsity of onset targets. Only $\psi$ is optimized in the decoupled configuration. The original activity model and its B2 probability cache are retained exactly.

\paragraph{Onset aggregation and event association.}
Onset learning uses recording-level feature windows of 450 frames and a stride of 150 frames. Let $\mathcal W_t^{\mathrm{valid}}$ denote onset windows covering frame $t$ at valid positions, with $u_t^{(w)}$ the logit at the corresponding local position. For nonempty $\mathcal W_t^{\mathrm{valid}}$, onset predictions are combined as probabilities,
\begin{equation}
 \bar r_t=\frac{1}{|\mathcal W_t^{\mathrm{valid}}|}\sum_{w\in\mathcal W_t^{\mathrm{valid}}}\sigma(u_t^{(w)}),
 \label{eq:onsetoverlap}
\end{equation}
which is distinct from the activity logit aggregation in Eq.~\ref{eq:overlap}. Uncovered recording frames have $\bar r_t=0$. Peaks above a validation-selected threshold are retained using temporal non-maximum suppression (NMS). A peak at recording index $t$ produces candidate time $\tau=\operatorname{time}(t)$.

Each candidate is associated with classes using the fixed activity probabilities. For time $\tau$, define the post- and pre-onset maxima
\begin{align}
 A_c^+(\tau)&=\max_{t:\operatorname{time}(t)\in[\tau+d_+,\tau+d_++0.16]}p_{t,c},\\
 A_c^-(\tau)&=\max_{t:\operatorname{time}(t)\in[\tau+d_--0.16,\tau+d_-)}p_{t,c},
\end{align}
where the maxima range over recording frames and $d_+,d_-$ are validation-selected evidence shifts in seconds. A class is accepted if
\begin{equation}
 A_c^+(\tau)\geq \eta_c,\qquad A_c^+(\tau)-0.5A_c^-(\tau)\geq 0.08.
 \label{eq:association}
\end{equation}
Here $\eta_c$ is the class-specific activity threshold. The second condition penalizes pre-existing activity while favoring candidates with strong post-onset evidence. Each accepted class yields $(\tau,c)$ at the original candidate time; multiple labels may be associated with one start. Duplicate same-class events within 1~ms are merged. The final direct configuration uses learned IPT peaks alone, without note-onset candidates. The note-onset baseline uses the same activity cache and association rule, providing a comparison of onset sources.

\section{Experimental Setup}
\label{sec:setup}
\subsection{Dataset and Protocol}
Experiments use Guzheng Tech99 \cite{li2023tech99}, comprising 99 continuous recordings with approximately 151 minutes of audio. The official recording-disjoint partition consists of 79 training, 10 validation, and 10 test recordings. The seven technique classes are \textit{chanyin} (vibrato), \textit{boxian} (plucking), \textit{shanghua} (upward slide), \textit{xiahua} (downward slide), a merged class containing \textit{huazhi/guazou/lianmo/liantuo}, \textit{yaozhi} (tremolo), and \textit{dianyin}. The merged class is retained as defined by the dataset. Annotations provide intervals for potentially overlapping techniques. The archived data contain 15,838 annotation rows before onset deduplication; the event scorer removes same-class duplicates within 1~ms. Training-event counts after deduplication range from 56 for \textit{yaozhi} to 9,063 for \textit{boxian} (Appendix~\ref{app:implementation}).

Earlier representation experiments use recording-disjoint five-fold cross-validation (CV) on the 89 training and validation recordings, covering clip-level probing and frame-level adaptation. Subsequent encoder adaptation and context-length experiments use the official 79/10/10 partition. Historical development results are documented in the appendix. Model configuration and checkpoint selection in the final continuation experiments are based on validation metrics.

\subsection{Implementation and Model Selection}
The context experiments initialize from a common task-adapted 12-block continuation checkpoint. AdamW updates the encoder and temporal head at learning rates of $3\times10^{-6}$ and $10^{-4}$, respectively, with weight decay 0.005 and gradient clipping at 1. Training targets approximately eight original 3-s segments per update, subject to grouping by context length, with a maximum of 24 epochs and patience of eight. Checkpoints are selected by validation frame macro-\Fone{} at threshold 0.5. A historical sweep with seed 42 considers eight context lengths. The selected 18-s configuration and the 3-s baseline are subsequently evaluated through paired continuation runs with seeds 42, 123, and 3407, using the common task-adapted initialization. B2 reuses the corresponding B1 checkpoint without additional training.

The decoupled onset head is initialized on the adapted 18-s representation and optimized using AdamW with learning rate $4\times10^{-4}$ and weight decay 0.007. Training employs rarity-weighted onset supervision and window sampling. Onset checkpoints and decoder settings are selected using validation event macro-\Fone{}; the class thresholds and evidence shifts are inherited from the note-onset baseline's validation-selected decoder. Detailed hyperparameters and the joint objective are specified in Appendix~\ref{app:implementation}.

\subsection{Comparison Systems}
The internal activity comparisons consist of the short-context baseline (B0; 3-s training and inference), the long-context model (B1; 18-s training and non-overlapping inference), and MuQ-LC (B2; the same B1 weights and overlapping inference). For event detection, the note-onset baseline combines B2 activity with candidates from an independently trained note model using frozen MuQ representations and separate temporal heads for note onset and pitch. The final decoupled IPT configuration replaces these candidates with learned technique starts while preserving the activity cache. A joint comparator initializes from the B1 model and trained onset head, then updates the last four encoder blocks, final normalization, and frame head alongside the onset head using activity supervision and teacher distillation. Its frame inference also uses a full-window readout and probability averaging, rather than B2's slice-wise readout and logit averaging. The event experiment's validation-based deployment ranking selects joint direct decoding. MuQ-LC adopts decoupled direct decoding to retain the original B2 activity output exactly, with the joint configuration serving as a comparator. Published results for the multi-scale network and MERTech, including the MERTech ablation configurations, provide benchmark comparisons on Tech99 \cite{li2024mertech}.

\subsection{Evaluation Metrics}
Following the Tech99 evaluation setting adopted by MERTech \cite{li2023tech99,li2024mertech} and standard MIR evaluation principles \cite{raffel2014mireval,mesaros2016metrics}, performance is measured by frame-level and event-level macro- and micro-\Fone{}. MuQ-LC scores activity on its native 25-Hz grid, whereas MERTech uses a 75-Hz feature grid and applies \texttt{mir\_eval} to decoded intervals and their reconstructed frame activity. For class $c$, $F_{1,c}=2TP_c/(2TP_c+FP_c+FN_c)$, with zero assigned when the denominator is zero. Counts are pooled across recordings before computing class scores. Macro-\Fone{} averages the seven class scores; micro-\Fone{} pools counts across all classes. Main activity experiments use threshold 0.5; event-system comparisons use class-specific thresholds selected on validation data. For events, a prediction matches a reference if the recording and technique class coincide and the absolute onset difference is at most 50~ms. Matching is one-to-one within each recording and class, with neither offset nor pitch included. All reported \Fone{} values are percentages; differences are percentage points (pp), computed before rounding. Three-seed results are summarized by the mean and sample standard deviation.

\section{Results and Analysis}
\label{sec:results}
\subsection{Overall Performance}
Table~\ref{tab:t01_overall} summarizes the activity and event results. MuQ-LC achieves 88.11\% frame macro-\Fone{} and 93.46\% frame micro-\Fone{} at a fixed activity threshold of 0.5. The decoupled onset configuration obtains 81.06\% event macro-\Fone{} and 92.74\% event micro-\Fone{}. The associated standard deviations are 0.26, 0.11, 0.24, and 0.12 percentage points, respectively. Together, these scores characterize the dense activity and localized event outputs under their respective threshold protocols.

MuQ-LC obtains higher scores than the published MERTech results on all four metrics. Relative to the full MERTech system, the differences are +7.71 and +3.46~pp in frame macro- and micro-\Fone{}, and +4.96 and +1.14~pp in event macro- and micro-\Fone{}, respectively. Compared with the strongest frame-level configurations reported in the MERTech ablation study, MuQ-LC further achieves gains of 3.61~pp over IPT+Pitch+Onset in frame macro-\Fone{} (84.50\%) and 1.76~pp over IPT+Pitch in frame micro-\Fone{} (91.70\%) \cite{li2024mertech}. The following analyses examine the contributions of representation adaptation, context extension, overlapping inference, and event localization.

\begin{table*}[t]
\centering
\caption{Performance on the official Tech99 split (\%). Multi-scale and MERTech scores are reported in MERTech (Tables 1--2) \cite{li2024mertech}. MuQ-LC scores are mean $\pm$ sample SD over three continuation seeds; frame threshold is 0.5 and event decoding uses validation-calibrated thresholds. PP: post-processing. Bold indicates the highest reported mean or point score.}
\label{tab:t01_overall}
\small
\begin{tabular}{lrrrr}
\toprule
System & Frame macro & Frame micro & Event macro & Event micro \\
\midrule
Multi-scale network & 69.30 & 86.50 & 38.90 & 38.80 \\
MERTech: IPT probing & 60.30 & 81.20 & 16.50 & 19.90 \\
MERTech: IPT fine-tune & 81.80 & 91.00 & 54.70 & 49.00 \\
MERTech: IPT + Pitch & 83.40 & 91.70 & 54.80 & 50.00 \\
MERTech: IPT + Pitch + Onset (no PP) & 84.50 & 91.40 & 57.00 & 50.20 \\
MERTech: full system (with PP) & 80.40 & 90.00 & 76.10 & 91.60 \\
\midrule
MuQ-LC + decoupled IPT & $\mathbf{88.11}\pm0.26$ & $\mathbf{93.46}\pm0.11$ & $\mathbf{81.06}\pm0.24$ & $\mathbf{92.74}\pm0.12$ \\
\bottomrule
\end{tabular}
\end{table*}

\subsection{Representation Adaptation}
\label{sec:adaptresults}
The initial analysis examines how representation layer selection and temporal readout affect the accessibility of technique information. Thirteen frozen MuQ representations, including the frontend output and 12 contextual layers, are evaluated using pooled 3-s clip representations. As shown in Table~\ref{tab:t02_adaptation}, the pooled fourth-layer representation achieves 73.76\% clip macro-\Fone{}, outperforming the pooled final-layer representation (68.11\%) by 5.65~pp. The layer-wise score distribution (Appendix~\ref{app:representation}) highlights the value of intermediate music representations for fine-grained technique recognition and is consistent with the task-dependent layer utility observed in prior music representation studies \cite{kanatas2026layer,zhou2025layerwise,bai2026dimension}.

Temporal readout substantially improves the recognition performance of frozen MuQ representations. With clip-level supervision, the last-layer temporal configuration reaches 77.78\% clip macro-\Fone{}, a gain of 9.67~pp over its pooled counterpart (68.11\%). Adding frame-level supervision further raises the score to 78.87\% (+1.09~pp), for a total gain of 10.76~pp. The best temporal configuration achieves 80.47\% (Appendix~\ref{app:representation}). These results demonstrate the complementary benefits of sequence-preserving readout and dense supervision, motivating their integration in MuQ-LC.

Dense supervision and encoder adaptation are examined separately in a frame-level five-fold development protocol. A frame-first model achieves 73.52\% macro-\Fone{}. Updating blocks 2--3 increases this value to 74.62\%, and extending the adapted range to blocks 2--5 yields 76.19\%. Adapting blocks 1--6 reaches 76.38\%, representing a gain of 2.86~pp over the frame-first baseline. The variation across adaptation ranges, including 76.07\% for blocks 2--6, demonstrates that recognition performance depends on both the extent and placement of trainable blocks. These improvements establish encoder adaptation as an effective component of dense technique recognition and motivate the subsequent investigation of full-encoder adaptation.

Following the development-set analysis, encoder adaptation is extended to all 12 contextual blocks. The resulting model achieves 82.67\% frame macro-\Fone{} on the official test split at a threshold of 0.5. The full-encoder continuation checkpoint provides a common task-adapted initialization for the subsequent context-length experiments, enabling a controlled investigation of temporal context within the established recognition architecture.

\begin{table}[t]
\centering
\caption{Selected adaptation experiments (\%). Clip-level development CV (top): five folds and three seeds. Frame-level development CV (bottom): five folds, seed 42.}
\label{tab:t02_adaptation}
\small
\begin{tabular}{lrr}
\toprule
Configuration & Macro & Micro \\
\midrule
Frozen pooled H4 & 73.76 & -- \\
Frozen pooled H12 & 68.11 & -- \\
Best temporal configuration & 80.47 & -- \\
Last-layer temporal & 78.87 & -- \\
\midrule
\multicolumn{3}{l}{\textit{Frame-level development CV (five folds, seed 42)}} \\
Frame-first & 73.52 & 86.80 \\
L2--L3 & 74.62 & 87.64 \\
L2--L4 & 75.03 & 87.92 \\
L2--L5 & 76.19 & 88.21 \\
L2--L6 & 76.07 & 88.13 \\
L1--L6 & 76.38 & 88.11 \\
\bottomrule
\end{tabular}
\end{table}

\subsection{Effect of Context Length}
\label{sec:contextresults}
\begin{figure}[t]
\centering
\includegraphics[width=\columnwidth]{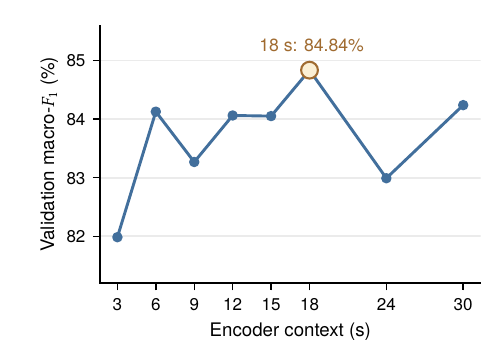}
\caption{Validation frame macro-\Fone{} for eight encoder context lengths in the historical single-seed sweep (seed 42, threshold 0.5). The 18-s setting was selected by validation macro-\Fone{}. Lines connect measured configurations.}
\Description{A line plot showing a nonmonotonic validation macro-F1 trend with the highest point at 18 seconds.}
\label{fig:sweep}
\end{figure}
The relationship between downstream encoder context and activity recognition is first evaluated across 3, 6, 9, 12, 15, 18, 24, and 30 seconds (Fig.~\ref{fig:sweep}). The historical single-seed experiment yields 81.98\% validation macro-\Fone{} for 3~s and 84.13\% for 6~s. Values for 12 and 15~s are both approximately 84.05\%, while the 18-s configuration reaches 84.84\%. Extending the context further does not produce a monotonic validation improvement: the 24-s and 30-s configurations achieve 82.99\% and 84.24\%, respectively.

The 18-s configuration achieves the highest validation macro-\Fone{} among the eight evaluated context lengths, improving on the 3-s setting by 2.85~pp. It is therefore selected for the subsequent three-seed paired experiments. The complete validation and test results are reported in Table~\ref{tab:ta01_context_sweep}. Complementing prior research on the multi-scale temporal characteristics of playing techniques \cite{li2023tech99,wang2026viotech}, this analysis establishes encoder context length as an important design parameter for continuous technique detection and provides an empirical basis for the long-context configuration of MuQ-LC.

\subsection{Long-Context and Overlap Ablation}
\begin{table*}[t]
\centering
\caption{Paired context and overlap ablation (F1, \%). Fixed threshold 0.5; mean $\pm$ sample SD across seeds 42, 123, and 3407. B1 and B2 share weights.}
\label{tab:t03_context_ablation}
\small
\begin{tabular}{lrrrr}
\toprule
Model & Val macro & Val micro & Test macro & Test micro \\
\midrule
Short context (3 s) & $82.06\pm0.15$ & $90.52\pm0.04$ & $82.89\pm0.94$ & $91.80\pm0.19$ \\
Long context (18 s) & $84.85\pm0.22$ & $91.80\pm0.34$ & $85.80\pm0.23$ & $93.01\pm0.15$ \\
MuQ-LC (18 s / 6 s) & $85.23\pm0.23$ & $92.38\pm0.19$ & $88.11\pm0.26$ & $93.46\pm0.11$ \\
\bottomrule
\end{tabular}
\end{table*}

Table~\ref{tab:t03_context_ablation} separates context extension from overlapping inference. Changing the encoder context from 3 to 18 seconds increases the test frame macro-\Fone{} from 82.89\% to 85.80\% (+2.91~pp), while the micro-\Fone{} rises from 91.80\% to 93.01\%. Validation macro-\Fone{} improves from 82.06\% to 84.85\%. The comparison uses the same frontend and head architecture, with a shared continuation checkpoint at initialization, and therefore examines the effect of longer training and inference context in this architecture.

Applying overlapping inference to the 18-s checkpoint yields a further test macro-\Fone{} gain of 2.30~pp and a micro-\Fone{} gain of 0.44~pp. The resulting MuQ-LC activity model reaches 88.11\% and 93.46\%, respectively. The validation macro-\Fone{} also increases to 85.23\%. Both stages improve macro-\Fone{} for each of the three paired seeds on validation and test. The total test macro-\Fone{} gain of 5.22~pp therefore reflects two distinct operations: adapting the encoder to longer feature sequences and aggregating multiple contextual estimates of each frame. Since B1 and B2 have identical weights, the latter contribution is attributable to the inference procedure rather than additional training. Alternative aggregation strategies are compared in Appendix~\ref{app:context}.

\begin{figure}[t]
\centering
\includegraphics[width=\columnwidth]{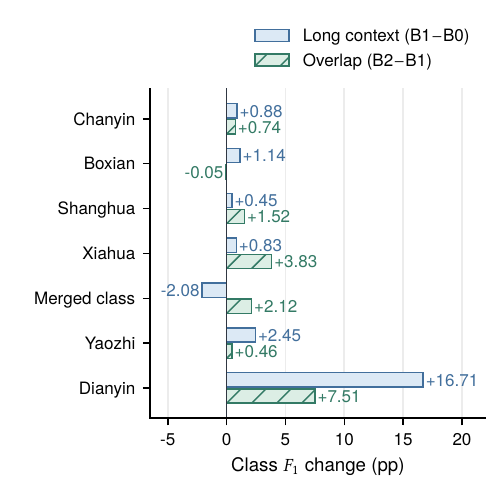}
\caption{Signed class-wise frame-\Fone{} changes from context extension (B1--B0) and overlapping inference (B2--B1), in pp. Values are differences of three-seed test means at threshold 0.5. The merged class contains huazhi/guazou/lianmo/liantuo.}
\Description{Grouped bars distinguish context and overlap gains for seven techniques; dianyin has the largest total gain.}
\label{fig:gains}
\end{figure}
The gains are heterogeneous across techniques (Fig.~\ref{fig:gains}). The largest improvement occurs for \textit{dianyin}: the class score rises from 43.83\% to 60.54\% with context extension, and to 68.04\% with overlapping inference. Its total improvement is 24.22~pp. The score for \textit{xiahua} increases by 4.66~pp overall, of which approximately 3.83~pp comes from overlapping inference. \textit{Yaozhi} improves by 2.91~pp, primarily through context extension, and \textit{chanyin} and \textit{shanghua} gain 1.63 and 1.97~pp, respectively. The merged class exhibits a different pattern: extending the context lowers its score by 2.08~pp, while overlapping inference recovers approximately 2.12~pp, leaving its overall score nearly unchanged. For \textit{boxian}, the overlap component is slightly negative despite an overall gain of 1.09~pp.

The class-wise results reveal complementary effects of context extension and overlapping inference. Context extension accounts for most of the improvements in \textit{dianyin} and \textit{yaozhi}, while overlapping inference contributes substantially to \textit{xiahua} recognition. The 24.22~pp gain for \textit{dianyin}, together with the improvement in \textit{yaozhi}, highlights the utility of contextual modeling across techniques with heterogeneous temporal characteristics. Together, these findings establish technique-dependent benefits of the two temporal modeling strategies across the evaluated classes. Full class-level scores appear in Appendix~\ref{app:results}.

\subsection{From Activity to Event Localization}
\paragraph{Candidate diagnostics.}
Accurate event localization requires both reliable technique activity estimates and precise occurrence-start candidates. With the B2 activity sequence, the note-onset baseline achieves 79.18\% event macro-\Fone{}. To isolate the contribution of candidate quality, three diagnostic interventions modify the note-onset candidate sequence using annotated reference onsets while retaining the B2 activity predictions and event association procedure. Reference candidates are annotated onset times after 8-ms temporal suppression. Within each recording, sorted note and reference candidates are matched greedily one-to-one within $\pm50$~ms using onset time alone. The deletion intervention removes unmatched note-onset candidates, while the addition intervention supplies missed reference starts and retains the existing candidates. The combined intervention performs both operations and preserves the original timing of matched candidates.

\begin{table}[t]
\centering
\caption{Event candidate diagnostics on the test split (F1, \%). GT interventions modify the candidate sequence using annotated reference onsets.}
\label{tab:t04_candidate_diagnostics}
\small
\begin{tabular}{lrr}
\toprule
Condition & Macro & Micro \\
\midrule
Note-onset baseline & 79.18 & 91.18 \\
GT delete & 84.24 & 92.90 \\
GT append & 82.43 & 91.48 \\
GT delete + append & 88.52 & 93.20 \\
\bottomrule
\end{tabular}
\end{table}

\begin{table*}[t]
\centering
\caption{Activity and event performance (F1, \%). Frame scores use validation-selected thresholds. Event scores are mean $\pm$ sample SD. Decoupled IPT preserves the original B2 activity cache; Joint had the highest validation event macro-F1.}
\label{tab:t05_event_localization}
\small
\begin{tabular}{lrrrrr}
\toprule
Method & Val event macro & Frame macro & Frame micro & Event macro & Event micro \\
\midrule
Note-onset baseline & 75.81 & 87.64 & 93.42 & $79.18\pm0.34$ & $91.18\pm0.09$ \\
Decoupled IPT onset & 83.06 & 87.64 & 93.42 & $81.06\pm0.24$ & $92.74\pm0.12$ \\
Joint IPT onset & 83.59 & 86.49 & 93.05 & $81.05\pm0.63$ & $92.76\pm0.13$ \\
\bottomrule
\end{tabular}
\end{table*}

Table~\ref{tab:t04_candidate_diagnostics} shows that deletion raises macro-\Fone{} to 84.24\% (+5.07~pp), while addition raises it to 82.43\% (+3.26~pp). The combined intervention reaches 88.52\%, a 9.34~pp improvement relative to the baseline, with micro-\Fone{} increasing to 93.20\%. The substantial gains obtained under fixed activity predictions and event association identify candidate precision and temporal coverage as key factors in event localization. These findings motivate the direct learning of playing technique onsets from task-adapted contextual representations, providing an event-specific temporal signal for the established activity model.

\paragraph{Decoupled and joint onset learning.}

The decoupled IPT configuration achieves 81.06\% test event macro-\Fone{} and 92.74\% micro-\Fone{}, improving on note-onset anchoring by 1.89 and 1.56~pp, respectively (Table~\ref{tab:t05_event_localization}). Validation event macro-\Fone{} increases from 75.81\% to 83.06\%, with improvements observed across all three seeds. The event decoder operates on the original B2 activity cache, preserving the activity predictions exactly. The corresponding frame macro- and micro-\Fone{} remain at 87.64\% and 93.42\% under validation-calibrated class thresholds, and at 88.11\% and 93.46\% under the fixed threshold of 0.5 (Table~\ref{tab:t03_context_ablation}). MERTech's onset-conditioned post-processing increases event macro-\Fone{} by 19.10~pp relative to its IPT+Pitch+Onset configuration, accompanied by a 4.10~pp reduction in frame macro-\Fone{} \cite{li2024mertech}. In MuQ-LC, replacing note-onset anchoring with independently learned IPT onsets improves event macro-\Fone{} by 1.89~pp while fully preserving the dense activity predictions.

\begin{figure}[t]
\centering
\includegraphics[width=\columnwidth]{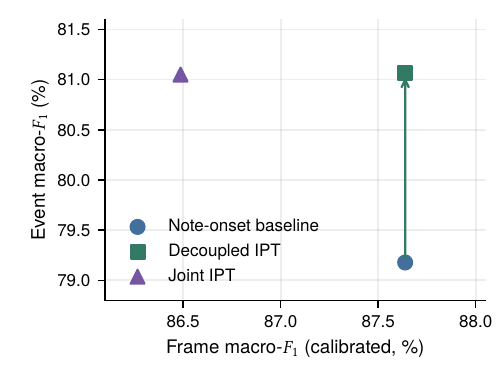}
\caption{Activity and event macro-\Fone{} for note-onset anchoring, decoupled IPT, and joint IPT (three-seed test means). Activity scores use validation-selected class thresholds. Joint adaptation also changes the frame readout and fusion procedure.}
\Description{A scatter plot showing decoupled onset learning at the same frame score as the note-onset baseline and a higher event score, with joint adaptation at a lower frame score and similar event score.}
\label{fig:tradeoff}
\end{figure}
Joint IPT adaptation achieves a comparable event macro-\Fone{} of 81.05\% and micro-\Fone{} of 92.76\%, with the highest validation event macro-\Fone{} (83.59\%). Its test frame macro-\Fone{} is 86.49\%, compared with 87.64\% for the fixed-cache configurations (Fig.~\ref{fig:tradeoff}). The decoupled configuration retains 1.15~pp more frame macro-\Fone{} than the joint training and inference configuration, which also changes readout scope and aggregation. With comparable event scores, independent onset optimization provides an activity-preserving solution for the combined task.

Class-wise event results further distinguish temporal coverage from class assignment. The event \Fone{} for \textit{yaozhi}, for example, increases from 37.12\% to 45.88\% with decoupled onset learning, while that for \textit{shanghua} rises from 90.87\% to 94.04\% and \textit{boxian} from 94.21\% to 95.94\%. The large frame score of \textit{yaozhi} and its substantially lower event score illustrate the different counting units: a sustained correctly classified region may still have a missed or additional predicted start. Candidate generation, class-specific event heads, and hybrid decoding were also examined during development; the results in Appendix~\ref{app:results} provide additional context for choosing a shared onset head with direct decoding. The activity-preserving comparison isolates the change in onset source at the level of the final decoded system.

\section{Conclusion}
MuQ-LC provides a staged approach to continuous guzheng playing technique detection using task-adapted pretrained music features. Representation experiments characterize the effects of layer choice, temporal readout, and encoder adaptation, motivating an activity model that retains local supervision while expanding contextual processing. The context sweep and paired continuation experiments demonstrate complementary gains from extending encoder context and combining overlapping logits. On Tech99, the resulting activity model reaches 88.11\% test frame macro-\Fone{}. A separately trained shared technique-onset head improves event macro-\Fone{} to 81.06\%, while preserving the established activity estimates exactly. The system scores above the published MERTech full system on all four evaluation metrics and exceeds its strongest frame-level ablation. The diagnostic interventions and joint-training comparison show the value of separating activity estimation from event localization while retaining the original activity sequence. These findings support treating representation adaptation, contextual inference, and event localization as related but separately examinable design dimensions in fine-grained music information retrieval.

\clearpage
\bibliographystyle{ACM-Reference-Format}
\bibliography{references}


\begin{thebibliography}{32}


\ifx \showCODEN    \undefined \def \showCODEN     #1{\unskip}     \fi
\ifx \showISBNx    \undefined \def \showISBNx     #1{\unskip}     \fi
\ifx \showISBNxiii \undefined \def \showISBNxiii  #1{\unskip}     \fi
\ifx \showISSN     \undefined \def \showISSN      #1{\unskip}     \fi
\ifx \showLCCN     \undefined \def \showLCCN      #1{\unskip}     \fi
\ifx \shownote     \undefined \def \shownote      #1{#1}          \fi
\ifx \showarticletitle \undefined \def \showarticletitle #1{#1}   \fi
\ifx \showURL      \undefined \def \showURL       {\relax}        \fi
\providecommand\bibfield[2]{#2}
\providecommand\bibinfo[2]{#2}
\providecommand\natexlab[1]{#1}
\providecommand\showeprint[2][]{arXiv:#2}

\bibitem[Bai et~al\mbox{.}(2026)]%
        {bai2026dimension}
\bibfield{author}{\bibinfo{person}{Xueying Bai}, \bibinfo{person}{Nanmu Hui},
  \bibinfo{person}{Yanping Li}, {and} \bibinfo{person}{Xiaowei Han}.}
  \bibinfo{year}{2026}\natexlab{}.
\newblock \showarticletitle{{Dimension-Wise Probing of Music Foundation Models
  for Chinese and Western Music: Timbre, Technique, Key, and Mode}}.
\newblock \bibinfo{journal}{\emph{Electronics}} \bibinfo{volume}{15},
  \bibinfo{number}{15} (\bibinfo{year}{2026}), \bibinfo{pages}{3421}.
\newblock
\href{https://doi.org/10.3390/electronics15153421}{doi:\nolinkurl{10.3390/electronics15153421}}


\bibitem[Chen et~al\mbox{.}(2015)]%
        {chen2015guitar}
\bibfield{author}{\bibinfo{person}{Yuan-Ping Chen}, \bibinfo{person}{Li Su},
  {and} \bibinfo{person}{Yi-Hsuan Yang}.} \bibinfo{year}{2015}\natexlab{}.
\newblock \showarticletitle{{Electric Guitar Playing Technique Detection in
  Real-World Recordings Based on F0 Sequence Pattern Recognition}}. In
  \bibinfo{booktitle}{\emph{Proceedings of the International Society for Music
  Information Retrieval Conference}}.
\newblock
\urldef\tempurl%
\url{https://archives.ismir.net/ismir2015/paper/000119.pdf}
\showURL{%
\tempurl}


\bibitem[Foscarin et~al\mbox{.}(2024)]%
        {foscarin2024beatthis}
\bibfield{author}{\bibinfo{person}{Francesco Foscarin}, \bibinfo{person}{Jan
  Schlüter}, {and} \bibinfo{person}{Gerhard Widmer}.}
  \bibinfo{year}{2024}\natexlab{}.
\newblock \showarticletitle{{Beat This! Accurate Beat Tracking Without DBN
  Postprocessing}}. In \bibinfo{booktitle}{\emph{Proceedings of the
  International Society for Music Information Retrieval Conference}}.
\newblock
\urldef\tempurl%
\url{https://arxiv.org/abs/2407.21658}
\showURL{%
\tempurl}


\bibitem[Gulati et~al\mbox{.}(2020)]%
        {gulati2020conformer}
\bibfield{author}{\bibinfo{person}{Anmol Gulati}, \bibinfo{person}{James Qin},
  \bibinfo{person}{Chung-Cheng Chiu}, \bibinfo{person}{Niki Parmar},
  \bibinfo{person}{Yu Zhang}, \bibinfo{person}{Jiahui Yu}, \bibinfo{person}{Wei
  Han}, \bibinfo{person}{Shibo Wang}, \bibinfo{person}{Zhengdong Zhang},
  \bibinfo{person}{Yonghui Wu}, {and} \bibinfo{person}{Ruoming Pang}.}
  \bibinfo{year}{2020}\natexlab{}.
\newblock \showarticletitle{{Conformer}: Convolution-augmented Transformer for
  Speech Recognition}. In \bibinfo{booktitle}{\emph{Interspeech 2020}}.
  \bibinfo{pages}{5036--5040}.
\newblock
\href{https://doi.org/10.21437/Interspeech.2020-3015}{doi:\nolinkurl{10.21437/Interspeech.2020-3015}}


\bibitem[Hawthorne et~al\mbox{.}(2018)]%
        {hawthorne2018onsets}
\bibfield{author}{\bibinfo{person}{Curtis Hawthorne}, \bibinfo{person}{Erich
  Elsen}, \bibinfo{person}{Jialin Song}, \bibinfo{person}{Adam Roberts},
  \bibinfo{person}{Ian Simon}, \bibinfo{person}{Colin Raffel},
  \bibinfo{person}{Jesse Engel}, \bibinfo{person}{Sageev Oore}, {and}
  \bibinfo{person}{Douglas Eck}.} \bibinfo{year}{2018}\natexlab{}.
\newblock \showarticletitle{{Onsets and Frames: Dual-Objective Piano
  Transcription}}. In \bibinfo{booktitle}{\emph{Proceedings of the
  International Society for Music Information Retrieval Conference}}.
\newblock
\urldef\tempurl%
\url{https://arxiv.org/abs/1710.11153}
\showURL{%
\tempurl}


\bibitem[Huang et~al\mbox{.}(2020)]%
        {huang2020guqin}
\bibfield{author}{\bibinfo{person}{Yu-Fen Huang}, \bibinfo{person}{Jeng-I
  Liang}, \bibinfo{person}{I-Chieh Wei}, {and} \bibinfo{person}{Li Su}.}
  \bibinfo{year}{2020}\natexlab{}.
\newblock \showarticletitle{{Joint Analysis of Mode and Playing Technique in
  Guqin Performance with Machine Learning}}. In
  \bibinfo{booktitle}{\emph{Proceedings of the International Society for Music
  Information Retrieval Conference}}.
\newblock
\href{https://doi.org/10.5281/zenodo.4245378}{doi:\nolinkurl{10.5281/zenodo.4245378}}


\bibitem[Kanatas et~al\mbox{.}(2026)]%
        {kanatas2026layer}
\bibfield{author}{\bibinfo{person}{Angelos-Nikolaos Kanatas},
  \bibinfo{person}{Yuexuan Kong}, \bibinfo{person}{Pablo Alonso-Jiménez},
  \bibinfo{person}{Xavier Serra}, {and} \bibinfo{person}{Dmitry Bogdanov}.}
  \bibinfo{year}{2026}\natexlab{}.
\newblock \showarticletitle{{What Makes a Good Layer? Assessing the Layer-Wise
  Intrinsic Properties of Music Foundation Models}}. In
  \bibinfo{booktitle}{\emph{Proceedings of the International Society for Music
  Information Retrieval Conference}}.
\newblock
\urldef\tempurl%
\url{https://arxiv.org/abs/2608.14819}
\showURL{%
\tempurl}


\bibitem[Kanatas et~al\mbox{.}(2025)]%
        {kanatas2025culturemert}
\bibfield{author}{\bibinfo{person}{Angelos-Nikolaos Kanatas},
  \bibinfo{person}{Charilaos Papaioannou}, {and} \bibinfo{person}{Alexandros
  Potamianos}.} \bibinfo{year}{2025}\natexlab{}.
\newblock \showarticletitle{{CultureMERT: Continual Pre-Training for
  Cross-Cultural Music Representation Learning}}. In
  \bibinfo{booktitle}{\emph{Proceedings of the International Society for Music
  Information Retrieval Conference}}.
\newblock
\urldef\tempurl%
\url{https://arxiv.org/abs/2506.17818}
\showURL{%
\tempurl}


\bibitem[Kong et~al\mbox{.}(2021)]%
        {kong2021highresolution}
\bibfield{author}{\bibinfo{person}{Qiuqiang Kong}, \bibinfo{person}{Bochen Li},
  \bibinfo{person}{Xuchen Song}, \bibinfo{person}{Yuan Wan}, {and}
  \bibinfo{person}{Yuxuan Wang}.} \bibinfo{year}{2021}\natexlab{}.
\newblock \showarticletitle{{High-Resolution Piano Transcription with Pedals by
  Regressing Onset and Offset Times}}.
\newblock \bibinfo{journal}{\emph{IEEE/ACM Transactions on Audio, Speech, and
  Language Processing}} (\bibinfo{year}{2021}).
\newblock
\urldef\tempurl%
\url{https://arxiv.org/abs/2010.01815}
\showURL{%
\tempurl}


\bibitem[Li et~al\mbox{.}(2023)]%
        {li2023tech99}
\bibfield{author}{\bibinfo{person}{Dichucheng Li}, \bibinfo{person}{Mingjin
  Che}, \bibinfo{person}{Wenwu Meng}, \bibinfo{person}{Yulun Wu},
  \bibinfo{person}{Yi Yu}, \bibinfo{person}{Fan Xia}, {and}
  \bibinfo{person}{Wei Li}.} \bibinfo{year}{2023}\natexlab{}.
\newblock \showarticletitle{{Frame-Level Multi-Label Playing Technique
  Detection Using Multi-Scale Network and Self-Attention Mechanism}}. In
  \bibinfo{booktitle}{\emph{IEEE International Conference on Acoustics, Speech
  and Signal Processing (ICASSP)}}.
\newblock
\urldef\tempurl%
\url{https://arxiv.org/abs/2303.13272}
\showURL{%
\tempurl}


\bibitem[Li et~al\mbox{.}(2024a)]%
        {li2024mertech}
\bibfield{author}{\bibinfo{person}{Dichucheng Li}, \bibinfo{person}{Yinghao
  Ma}, \bibinfo{person}{Weixing Wei}, \bibinfo{person}{Qiuqiang Kong},
  \bibinfo{person}{Yulun Wu}, \bibinfo{person}{Mingjin Che},
  \bibinfo{person}{Fan Xia}, \bibinfo{person}{Emmanouil Benetos}, {and}
  \bibinfo{person}{Wei Li}.} \bibinfo{year}{2024}\natexlab{a}.
\newblock \showarticletitle{{MERTech: Instrument Playing Technique Detection
  Using Self-Supervised Pretrained Model with Multi-Task Finetuning}}. In
  \bibinfo{booktitle}{\emph{IEEE International Conference on Acoustics, Speech
  and Signal Processing (ICASSP)}}.
\newblock
\urldef\tempurl%
\url{https://arxiv.org/abs/2310.09853}
\showURL{%
\tempurl}


\bibitem[Li et~al\mbox{.}(2022)]%
        {li2022guzheng}
\bibfield{author}{\bibinfo{person}{Dichucheng Li}, \bibinfo{person}{Yulun Wu},
  \bibinfo{person}{Qinyu Li}, \bibinfo{person}{Jiahao Zhao},
  \bibinfo{person}{Yi Yu}, \bibinfo{person}{Fan Xia}, {and}
  \bibinfo{person}{Wei Li}.} \bibinfo{year}{2022}\natexlab{}.
\newblock \showarticletitle{{Playing Technique Detection by Fusing Note Onset
  Information in Guzheng Performance}}. In
  \bibinfo{booktitle}{\emph{Proceedings of the International Society for Music
  Information Retrieval Conference}}.
\newblock
\urldef\tempurl%
\url{https://archives.ismir.net/ismir2022/paper/000037.pdf}
\showURL{%
\tempurl}


\bibitem[Li et~al\mbox{.}(2024b)]%
        {li2024mert}
\bibfield{author}{\bibinfo{person}{Yizhi Li}, \bibinfo{person}{Ruibin Yuan},
  \bibinfo{person}{Ge Zhang}, \bibinfo{person}{Yinghao Ma},
  \bibinfo{person}{Xingran Chen}, \bibinfo{person}{Hanzhi Yin},
  \bibinfo{person}{Chenghao Xiao}, \bibinfo{person}{Chenghua Lin},
  \bibinfo{person}{Anton Ragni}, \bibinfo{person}{Emmanouil Benetos},
  \bibinfo{person}{Norbert Gyenge}, \bibinfo{person}{Roger Dannenberg},
  \bibinfo{person}{Ruibo Liu}, \bibinfo{person}{Wenhu Chen},
  \bibinfo{person}{Gus Xia}, \bibinfo{person}{Yemin Shi},
  \bibinfo{person}{Wenhao Huang}, \bibinfo{person}{Zili Wang},
  \bibinfo{person}{Yike Guo}, {and} \bibinfo{person}{Jie Fu}.}
  \bibinfo{year}{2024}\natexlab{b}.
\newblock \showarticletitle{{MERT: Acoustic Music Understanding Model with
  Large-Scale Self-Supervised Training}}. In
  \bibinfo{booktitle}{\emph{International Conference on Learning
  Representations}}.
\newblock
\urldef\tempurl%
\url{https://arxiv.org/abs/2306.00107}
\showURL{%
\tempurl}


\bibitem[McCallum et~al\mbox{.}(2022)]%
        {mccallum2022representations}
\bibfield{author}{\bibinfo{person}{Matthew~C. McCallum}, \bibinfo{person}{Filip
  Korzeniowski}, \bibinfo{person}{Sergio Oramas}, \bibinfo{person}{Fabien
  Gouyon}, {and} \bibinfo{person}{Andreas~F. Ehmann}.}
  \bibinfo{year}{2022}\natexlab{}.
\newblock \showarticletitle{{Supervised and Unsupervised Learning of Audio
  Representations for Music Understanding}}. In
  \bibinfo{booktitle}{\emph{Proceedings of the International Society for Music
  Information Retrieval Conference}}.
\newblock
\urldef\tempurl%
\url{https://arxiv.org/abs/2210.03799}
\showURL{%
\tempurl}


\bibitem[Mesaros et~al\mbox{.}(2016)]%
        {mesaros2016metrics}
\bibfield{author}{\bibinfo{person}{Annamaria Mesaros}, \bibinfo{person}{Toni
  Heittola}, {and} \bibinfo{person}{Tuomas Virtanen}.}
  \bibinfo{year}{2016}\natexlab{}.
\newblock \showarticletitle{{Metrics for Polyphonic Sound Event Detection}}.
\newblock \bibinfo{journal}{\emph{Applied Sciences}} \bibinfo{volume}{6},
  \bibinfo{number}{6} (\bibinfo{year}{2016}), \bibinfo{pages}{162}.
\newblock
\href{https://doi.org/10.3390/app6060162}{doi:\nolinkurl{10.3390/app6060162}}


\bibitem[Mohseni et~al\mbox{.}(2026)]%
        {mohseni2026iranian}
\bibfield{author}{\bibinfo{person}{Pouya Mohseni}, \bibinfo{person}{Bagher
  BabaAli}, {and} \bibinfo{person}{Ichiro Fujinaga}.}
  \bibinfo{year}{2026}\natexlab{}.
\newblock \showarticletitle{{Adapting Music Foundation Models to Iranian
  Classical Music}}. In \bibinfo{booktitle}{\emph{Proceedings of the
  International Society for Music Information Retrieval Conference}}.
\newblock
\urldef\tempurl%
\url{https://ismir2026program.ismir.net/poster_448.html}
\showURL{%
\tempurl}


\bibitem[Papaioannou et~al\mbox{.}(2025)]%
        {papaioannou2025universal}
\bibfield{author}{\bibinfo{person}{Charilaos Papaioannou},
  \bibinfo{person}{Emmanouil Benetos}, {and} \bibinfo{person}{Alexandros
  Potamianos}.} \bibinfo{year}{2025}\natexlab{}.
\newblock \showarticletitle{{Universal Music Representations? Evaluating
  Foundation Models on World Music Corpora}}. In
  \bibinfo{booktitle}{\emph{Proceedings of the International Society for Music
  Information Retrieval Conference}}.
\newblock
\urldef\tempurl%
\url{https://arxiv.org/abs/2506.17055}
\showURL{%
\tempurl}


\bibitem[Raffel et~al\mbox{.}(2014)]%
        {raffel2014mireval}
\bibfield{author}{\bibinfo{person}{Colin Raffel}, \bibinfo{person}{Brian
  McFee}, \bibinfo{person}{Eric~J. Humphrey}, \bibinfo{person}{Justin Salamon},
  \bibinfo{person}{Oriol Nieto}, \bibinfo{person}{Dawen Liang}, {and}
  \bibinfo{person}{Daniel P.~W. Ellis}.} \bibinfo{year}{2014}\natexlab{}.
\newblock \showarticletitle{{mir\_eval: A Transparent Implementation of Common
  MIR Metrics}}. In \bibinfo{booktitle}{\emph{Proceedings of the International
  Society for Music Information Retrieval Conference}}.
\newblock
\urldef\tempurl%
\url{https://archives.ismir.net/ismir2014/paper/000320.pdf}
\showURL{%
\tempurl}


\bibitem[Stefani and Turchet(2025)]%
        {stefani2025guitar}
\bibfield{author}{\bibinfo{person}{Domenico Stefani} {and}
  \bibinfo{person}{Luca Turchet}.} \bibinfo{year}{2025}\natexlab{}.
\newblock \showarticletitle{{Real-Time Playing Technique Recognition Embedded
  in a Smart Acoustic Guitar}}.
\newblock \bibinfo{journal}{\emph{EURASIP Journal on Audio, Speech, and Music
  Processing}}  \bibinfo{volume}{2025} (\bibinfo{year}{2025}),
  \bibinfo{pages}{28}.
\newblock
\href{https://doi.org/10.1186/s13636-025-00413-6}{doi:\nolinkurl{10.1186/s13636-025-00413-6}}


\bibitem[Su et~al\mbox{.}(2019)]%
        {su2019tent}
\bibfield{author}{\bibinfo{person}{Ting-Wei Su}, \bibinfo{person}{Yuan-Ping
  Chen}, \bibinfo{person}{Li Su}, {and} \bibinfo{person}{Yi-Hsuan Yang}.}
  \bibinfo{year}{2019}\natexlab{}.
\newblock \showarticletitle{{TENT: Technique-Embedded Note Tracking for
  Real-World Guitar Solo Recordings}}.
\newblock \bibinfo{journal}{\emph{Transactions of the International Society for
  Music Information Retrieval}} (\bibinfo{year}{2019}).
\newblock
\href{https://doi.org/10.5334/tismir.23}{doi:\nolinkurl{10.5334/tismir.23}}


\bibitem[Toyama et~al\mbox{.}(2023)]%
        {toyama2023hft}
\bibfield{author}{\bibinfo{person}{Keisuke Toyama}, \bibinfo{person}{Taketo
  Akama}, \bibinfo{person}{Yukara Ikemiya}, \bibinfo{person}{Yuhta Takida},
  \bibinfo{person}{Wei-Hsiang Liao}, {and} \bibinfo{person}{Yuki Mitsufuji}.}
  \bibinfo{year}{2023}\natexlab{}.
\newblock \showarticletitle{{Automatic Piano Transcription with Hierarchical
  Frequency-Time Transformer}}. In \bibinfo{booktitle}{\emph{Proceedings of the
  International Society for Music Information Retrieval Conference}}.
\newblock
\urldef\tempurl%
\url{https://arxiv.org/abs/2307.04305}
\showURL{%
\tempurl}


\bibitem[Wang et~al\mbox{.}(2026a)]%
        {wang2026viotech}
\bibfield{author}{\bibinfo{person}{Ting-Kang Wang}, \bibinfo{person}{Vincent
  K.~M. Cheung}, \bibinfo{person}{Nan-Tien Lai}, {and} \bibinfo{person}{Li
  Su}.} \bibinfo{year}{2026}\natexlab{a}.
\newblock \showarticletitle{{VioTech: Frame-Level Violin Playing Technique
  Detection with Multi-Scale Mixture-of-Experts}}. In
  \bibinfo{booktitle}{\emph{Proceedings of the International Society for Music
  Information Retrieval Conference}}.
\newblock
\urldef\tempurl%
\url{https://arxiv.org/abs/2611.02680}
\showURL{%
\tempurl}


\bibitem[Wang et~al\mbox{.}(2026b)]%
        {wang2026vioptt}
\bibfield{author}{\bibinfo{person}{Ting-Kang Wang}, \bibinfo{person}{Yueh-Po
  Peng}, \bibinfo{person}{Li Su}, {and} \bibinfo{person}{Vincent K.~M.
  Cheung}.} \bibinfo{year}{2026}\natexlab{b}.
\newblock \showarticletitle{{VioPTT: Violin Technique-Aware Transcription from
  Synthetic Data Augmentation}}. In \bibinfo{booktitle}{\emph{IEEE
  International Conference on Acoustics, Speech and Signal Processing
  (ICASSP)}}.
\newblock
\urldef\tempurl%
\url{https://arxiv.org/abs/2509.23759}
\showURL{%
\tempurl}


\bibitem[Wang et~al\mbox{.}(2019)]%
        {wang2019erhu}
\bibfield{author}{\bibinfo{person}{Zehao Wang}, \bibinfo{person}{Jingru Li},
  \bibinfo{person}{Xiaoou Chen}, \bibinfo{person}{Zijin Li},
  \bibinfo{person}{Shicheng Zhang}, \bibinfo{person}{Baoqiang Han}, {and}
  \bibinfo{person}{Deshun Yang}.} \bibinfo{year}{2019}\natexlab{}.
\newblock \bibinfo{title}{{Musical Instrument Playing Technique Detection Based
  on FCN: Using Chinese Bowed-Stringed Instrument as an Example}}.
\newblock \bibinfo{howpublished}{arXiv:1910.09021}.
\newblock
\urldef\tempurl%
\url{https://arxiv.org/abs/1910.09021}
\showURL{%
\tempurl}


\bibitem[Wei and Yoshii(2025)]%
        {wei2025sparse}
\bibfield{author}{\bibinfo{person}{Weixing Wei} {and}
  \bibinfo{person}{Kazuyoshi Yoshii}.} \bibinfo{year}{2025}\natexlab{}.
\newblock \showarticletitle{{Efficient Transformer-Based Piano Transcription
  with Sparse Attention Mechanisms}}. In \bibinfo{booktitle}{\emph{Asia Pacific
  Signal and Information Processing Association Annual Summit and Conference
  (APSIPA ASC)}}.
\newblock
\urldef\tempurl%
\url{https://www.apsipa.org/proceedings/2025/papers/APSIPA2025_P356.pdf}
\showURL{%
\tempurl}


\bibitem[Won et~al\mbox{.}(2024)]%
        {won2024musicfm}
\bibfield{author}{\bibinfo{person}{Minz Won}, \bibinfo{person}{Yun-Ning Hung},
  {and} \bibinfo{person}{Duc Le}.} \bibinfo{year}{2024}\natexlab{}.
\newblock \showarticletitle{{A Foundation Model for Music Informatics}}. In
  \bibinfo{booktitle}{\emph{IEEE International Conference on Acoustics, Speech
  and Signal Processing (ICASSP)}}.
\newblock
\urldef\tempurl%
\url{https://arxiv.org/abs/2311.03318}
\showURL{%
\tempurl}


\bibitem[Yan et~al\mbox{.}(2021)]%
        {yan2021semicrf}
\bibfield{author}{\bibinfo{person}{Yujia Yan}, \bibinfo{person}{Frank
  Cwitkowitz}, {and} \bibinfo{person}{Zhiyao Duan}.}
  \bibinfo{year}{2021}\natexlab{}.
\newblock \showarticletitle{{Skipping the Frame-Level: Event-Based Piano
  Transcription with Neural Semi-CRFs}}. In \bibinfo{booktitle}{\emph{Advances
  in Neural Information Processing Systems}}.
\newblock
\urldef\tempurl%
\url{https://proceedings.neurips.cc/paper_files/paper/2021/file/ac53fab47b547a0d47b77e424cf119ba-Paper.pdf}
\showURL{%
\tempurl}


\bibitem[Yan and Duan(2024)]%
        {yan2024intervals}
\bibfield{author}{\bibinfo{person}{Yujia Yan} {and} \bibinfo{person}{Zhiyao
  Duan}.} \bibinfo{year}{2024}\natexlab{}.
\newblock \showarticletitle{{Scoring Time Intervals Using Non-Hierarchical
  Transformer for Automatic Piano Transcription}}. In
  \bibinfo{booktitle}{\emph{Proceedings of the International Society for Music
  Information Retrieval Conference}}.
\newblock
\urldef\tempurl%
\url{https://arxiv.org/abs/2404.09466}
\showURL{%
\tempurl}


\bibitem[Yuan et~al\mbox{.}(2023)]%
        {yuan2023marble}
\bibfield{author}{\bibinfo{person}{Ruibin Yuan}, \bibinfo{person}{Yinghao Ma},
  \bibinfo{person}{Yizhi Li}, \bibinfo{person}{Ge Zhang},
  \bibinfo{person}{Xingran Chen}, \bibinfo{person}{Hanzhi Yin},
  \bibinfo{person}{Le Zhuo}, \bibinfo{person}{Yiqi Liu},
  \bibinfo{person}{Jiawen Huang}, \bibinfo{person}{Zeyue Tian},
  \bibinfo{person}{Binyue Deng}, \bibinfo{person}{Ningzhi Wang},
  \bibinfo{person}{Chenghua Lin}, \bibinfo{person}{Emmanouil Benetos},
  \bibinfo{person}{Anton Ragni}, \bibinfo{person}{Norbert Gyenge},
  \bibinfo{person}{Roger Dannenberg}, \bibinfo{person}{Wenhu Chen},
  \bibinfo{person}{Gus Xia}, \bibinfo{person}{Wei Xue}, \bibinfo{person}{Si
  Liu}, \bibinfo{person}{Shi Wang}, \bibinfo{person}{Ruibo Liu},
  \bibinfo{person}{Yike Guo}, {and} \bibinfo{person}{Jie Fu}.}
  \bibinfo{year}{2023}\natexlab{}.
\newblock \showarticletitle{{MARBLE: Music Audio Representation Benchmark for
  Universal Evaluation}}. In \bibinfo{booktitle}{\emph{Advances in Neural
  Information Processing Systems}}.
\newblock
\urldef\tempurl%
\url{https://arxiv.org/abs/2306.10548}
\showURL{%
\tempurl}


\bibitem[Zhao et~al\mbox{.}(2022)]%
        {zhao2022beattransformer}
\bibfield{author}{\bibinfo{person}{Jingwei Zhao}, \bibinfo{person}{Gus Xia},
  {and} \bibinfo{person}{Ye Wang}.} \bibinfo{year}{2022}\natexlab{}.
\newblock \showarticletitle{{Beat Transformer: Demixed Beat and Downbeat
  Tracking with Dilated Self-Attention}}. In
  \bibinfo{booktitle}{\emph{Proceedings of the International Society for Music
  Information Retrieval Conference}}.
\newblock
\urldef\tempurl%
\url{https://arxiv.org/abs/2209.07140}
\showURL{%
\tempurl}


\bibitem[Zhou et~al\mbox{.}(2025)]%
        {zhou2025layerwise}
\bibfield{author}{\bibinfo{person}{Yizhi Zhou}, \bibinfo{person}{Haina Zhu},
  {and} \bibinfo{person}{Hangting Chen}.} \bibinfo{year}{2025}\natexlab{}.
\newblock \bibinfo{title}{{Layer-wise Investigation of Large-Scale
  Self-Supervised Music Representation Models}}.
\newblock \bibinfo{howpublished}{arXiv:2505.16306}.
\newblock
\urldef\tempurl%
\url{https://arxiv.org/abs/2505.16306}
\showURL{%
\tempurl}


\bibitem[Zhu et~al\mbox{.}(2025)]%
        {zhu2025muq}
\bibfield{author}{\bibinfo{person}{Haina Zhu}, \bibinfo{person}{Yizhi Zhou},
  \bibinfo{person}{Hangting Chen}, \bibinfo{person}{Jianwei Yu},
  \bibinfo{person}{Ziyang Ma}, \bibinfo{person}{Rongzhi Gu},
  \bibinfo{person}{Yi Luo}, \bibinfo{person}{Wei Tan}, {and}
  \bibinfo{person}{Xie Chen}.} \bibinfo{year}{2025}\natexlab{}.
\newblock \showarticletitle{{MuQ: Self-Supervised Music Representation Learning
  with Mel Residual Vector Quantization}}.
\newblock \bibinfo{journal}{\emph{IEEE Transactions on Audio, Speech and
  Language Processing}} (\bibinfo{year}{2025}).
\newblock
\urldef\tempurl%
\url{https://ieeexplore.ieee.org/document/11134793/}
\showURL{%
\tempurl}


\end{thebibliography}

\clearpage
\appendix
\raggedbottom
\setlength{\intextsep}{8pt plus 1pt minus 1pt}
\setlength{\floatsep}{6pt plus 1pt minus 1pt}
\setlength{\textfloatsep}{8pt plus 1pt minus 1pt}
\section{Representation and Adaptation Experiments}
\label{app:representation}
The preliminary studies examine alternative representations and adaptation choices before the final continuation experiments. An initial 3-s clip-level multi-label study compares pooled frozen MuQ features with explicit acoustic-mechanism features capturing onset, temporal, dynamic, and pitch-modulation descriptors. The report-level averages across five seeds are 64.06/83.31\% macro/micro-\Fone{} for MuQ, 54.31/75.08\% for the mechanism features, and 68.97/85.23\% for their fusion. The fusion improves macro-\Fone{} by 4.91~pp over pooled MuQ, demonstrating the complementary value of explicit acoustic information. These results motivate the subsequent investigations of representation layers, temporal readout, and dense supervision.

A subsequent recording-disjoint clip-level CV study evaluates 13 frozen layer representations and temporal readout variants. Table~\ref{tab:app_probes} records the mean clip macro-\Fone{} across five folds and three seeds. Table~\ref{tab:app_adaptation} reports selected five-fold frame-level adaptation experiments, performed with seed 42 on the development partition. These adaptation experiments examine trainable encoder block ranges, layer fusion, and auxiliary training objectives.

\begin{table}[H]
\centering
\caption{Frozen clip probes and temporal readouts (macro-\Fone{}, \%). Five development folds and three seeds.}
\label{tab:app_probes}
\small
\setlength{\tabcolsep}{3.5pt}
\begin{tabular}{lr}
\toprule
Representation & Macro \\
\midrule
Pooled H0 & 70.54 \\
Pooled H1 & 70.66 \\
Pooled H2 & 73.40 \\
Pooled H3 & 72.69 \\
Pooled H4 & 73.76 \\
Pooled H5 & 71.78 \\
Pooled H6 & 71.89 \\
Pooled H7 & 70.74 \\
Pooled H8 & 68.91 \\
Pooled H9 & 69.68 \\
Pooled H10 & 69.45 \\
Pooled H11 & 68.92 \\
Pooled H12 & 68.11 \\
Best temporal & 80.47 \\
Last-layer temporal & 78.87 \\
Last-layer clip-supervised & 77.78 \\
Temporal layer mixture & 80.09 \\
\bottomrule
\end{tabular}
\end{table}

\begin{table}[!htbp]
\centering
\caption{Frame-level adaptation on the development partition (five folds, seed 42; \%). Archive variants retain their original objectives.}
\label{tab:app_adaptation}
\small
\setlength{\tabcolsep}{3.5pt}
\begin{tabular}{lrr}
\toprule
Configuration & Macro & Micro \\
\midrule
Frame-first & 73.52 & 86.80 \\
Blocks 2--6, partial & 76.07 & 88.13 \\
Blocks 2--6, affine fusion & 76.30 & 88.19 \\
Blocks 2--6, softmax fusion & 75.50 & 87.95 \\
Blocks 2--3 & 74.62 & 87.64 \\
Blocks 2--4 & 75.03 & 87.92 \\
Blocks 2--5 & 76.19 & 88.21 \\
Blocks 3--6 & 74.89 & 87.04 \\
Blocks 4--6 & 73.99 & 87.10 \\
Blocks 1--3 & 75.89 & 88.59 \\
Blocks 1--6 & 76.38 & 88.11 \\
Joint auxiliary (B) & 76.11 & 88.11 \\
Joint pitch (C) & 75.72 & 87.93 \\
Joint pitch + onset (D) & 76.12 & 88.34 \\
Joint onset (E) & 76.63 & 88.41 \\
Layer 2; frozen & 73.18 & 86.46 \\
Layer 2; lora & 73.75 & 86.85 \\
Layer 2; partial & 73.67 & 87.03 \\
Layer 6; frozen & 72.95 & 85.86 \\
Layer 6; lora & 73.13 & 85.97 \\
Layer 6; partial & 73.47 & 86.11 \\
\bottomrule
\end{tabular}
\end{table}

\section{Context Selection and Aggregation}
\label{app:context}
Table~\ref{tab:ta01_context_sweep} lists the historical eight-length sweep, including the checkpoint epoch selected on validation macro-\Fone{} and the corresponding test scores (V/T: validation/test; Ma/Mi: macro/micro). The sweep uses seed 42, with the 18-s model subsequently used for the paired continuation study. Alternative inference readouts on that seed compare non-overlapping inference, nearest-center selection, uniform overlapping-logit averaging, and raised-cosine center-weighted aggregation (Table~\ref{tab:app_overlap}). Uniform averaging is selected by validation macro-\Fone{}. These alternatives use the same checkpoint and do not require retraining.

\begin{table}[!htbp]
\centering
\caption{Historical context-length sweep (F1, \%), seed 42, threshold 0.5. The 18-s configuration was selected by validation macro-F1 for the subsequent paired experiments.}
\label{tab:ta01_context_sweep}
\small
\setlength{\tabcolsep}{3.5pt}
\begin{tabular}{lrrrrr}
\toprule
Sec. & Epoch & V-Ma & V-Mi & T-Ma & T-Mi \\
\midrule
3 & 19 & 81.98 & 90.48 & 82.35 & 91.74 \\
6 & 4 & 84.13 & 90.58 & 82.14 & 91.06 \\
9 & 2 & 83.27 & 90.92 & 84.03 & 91.90 \\
12 & 3 & 84.06 & 91.18 & 85.81 & 92.15 \\
15 & 21 & 84.05 & 91.81 & 86.32 & 92.60 \\
18 & 22 & 84.84 & 91.85 & 85.89 & 93.08 \\
24 & 10 & 82.99 & 91.52 & 86.78 & 92.97 \\
30 & 18 & 84.24 & 91.85 & 86.43 & 92.64 \\
\bottomrule
\end{tabular}
\end{table}

\begin{table}[!htbp]
\centering
\caption{Inference aggregation comparison, seed 42; validation frame scores at threshold 0.5 (\%).}
\label{tab:app_overlap}
\small
\setlength{\tabcolsep}{3.5pt}
\begin{tabular}{lrr}
\toprule
Readout & Macro & Micro \\
\midrule
Non-overlap & 84.84 & 91.85 \\
Nearest center & 84.00 & 92.05 \\
Uniform logit mean & 85.05 & 92.37 \\
Center-weighted logits & 84.21 & 92.21 \\
\bottomrule
\end{tabular}
\end{table}

\section{Implementation Details}
\label{app:implementation}
\paragraph{Activity model.}
The frontend cache consists of 75-frame features per 3-s chunk, with recording and timestamp metadata. Consecutive chunks are grouped within recordings; a gap deviating from the expected 3-s succession by more than 55~ms breaks a chain. A shortened final group is processed with the available chunks. Individual invalid frames in cached chunks are masked in the head and loss; the contextual encoder uses the vendor block calls without an explicit attention mask. Group batches contain approximately eight original chunks, with the group batch size set to $\max(1,\lfloor 8/K\rfloor)$. Training uses AdamW, the learning rates specified in Sec.~\ref{sec:setup}, and clip-loss weight 0.25. Table~\ref{tab:app_epochs} lists the selected checkpoint epochs for the final six continuation runs.

\paragraph{Onset head and sampling.}
Each residual block contains a depthwise kernel-3 temporal convolution with the corresponding dilation, a pointwise channel-mixing convolution, group normalization with eight groups, GELU, dropout (0.15), and residual GELU. The projection before the residual stack also uses group normalization and GELU. Training event counts in class order are 1,604, 9,063, 563, 174, 555, 56, and 160, giving a median positive count of 555. For count $n_c$, the rarity weight is
\begin{equation}
 \rho_c=\operatorname{clip}\left(\sqrt{\frac{555}{\max(n_c,1)}},1,8\right).
\end{equation}
The local weight $\rho_\nu$ in Eq.~\ref{eq:iptloss} is the maximum of one and the class rarity weights $\rho_c$ of the onset supports covering the corresponding recording frame. Onset windows are sampled with weight $1+2I_{\mathrm{yaozhi}}+0.7I_{\mathrm{merged}}+0.4I_{\mathrm{dianyin}}$, where each indicator denotes the presence of a start of the corresponding technique. The frozen head uses a batch size of two, gradient clipping at 1, up to 24 epochs and patience of five. Windows with padded or uncovered positions exclude those positions from the loss and probability aggregation. Validation is performed at epoch 1, every second epoch, and the final scheduled epoch. At each evaluation, direct and hybrid decoding are compared over onset thresholds 0.15--0.85 (step 0.10) and NMS intervals of 80 and 160~ms. Checkpoints are ranked by validation event macro-\Fone{}, prioritizing those whose validation frame macro-\Fone{} is no more than 1.0~pp below the baseline. All selected checkpoints satisfy this frame guard. The selected epochs for seeds 42, 123, and 3407 are $(4,2,2)$ for frozen and $(2,8,1)$ for joint. All six select direct decoding with 80-ms NMS; the frozen thresholds are $(0.65,0.65,0.75)$ and the joint thresholds are $(0.45,0.35,0.75)$.

\paragraph{Association and joint comparator.}
The evidence intervals in Eq.~\ref{eq:association} have length 160~ms. An empty pre-onset interval contributes zero; an empty post-onset interval uses the nearest frame within 60~ms if available, otherwise the candidate is skipped. Thresholds for activity are selected per class on validation over 0.05--0.95 in increments of 0.025, with ties resolved by proximity to 0.5 and then by the higher threshold. The original note-onset decoder selects the post shift from $\{0,40,80\}$~ms and then the pre shift from $\{0,20,40,60\}$~ms on validation. Candidate timestamps are not shifted during association.

The joint comparator initializes from the adapted B1 checkpoint and the selected frozen onset head. It unfreezes blocks 9--12, final encoder normalization, and the activity head. The activity head operates over the full 450-frame window for both the joint model and its B1 teacher. With $\nu$ enumerating window-frame positions in the mini-batch as in Eq.~\ref{eq:iptloss}, $z_{\nu,c}$ denotes the joint full-window logit, $y_{\nu,c}$ its mapped recording-level target from Eq.~\ref{eq:frame_target}, and $v_\nu$ its validity mask. The joint objective is
\begin{align}
 \mathcal L_{\mathrm{joint}} &= \mathcal L_{\mathrm{IPT}}+0.3\mathcal L_f^{\mathrm{joint}}+\mathcal L_{\mathrm{KD}},\\
 \mathcal L_f^{\mathrm{joint}} &= \frac{\sum_{\nu,c}v_\nu b(z_{\nu,c},y_{\nu,c};a_c^f)}{\max(1,7\sum_\nu v_\nu)},\\
 \mathcal L_{\mathrm{KD}} &= \frac{\sum_{\nu,c}v_\nu b(z_{\nu,c},\tilde p_{\nu,c};1)}{\max(1,7\sum_\nu v_\nu)},
\end{align}
Here $a_c^f$ follows Eq.~\ref{eq:activity_weights}, counting each valid training recording frame once. The detached B1 teacher probability $\tilde p_{\nu,c}$ is computed on the same window and clipped to $[10^{-5},1-10^{-5}]$. Encoder and activity-head learning rates are $8\times10^{-6}$ and $2\times10^{-5}$, respectively. Unlike the original B2 inference path, joint inference applies the activity head to the full encoder window and averages activity probabilities. The comparison thus characterizes two complete optimization and inference configurations. The archived event experiment ranks eligible modes by mean validation event macro-\Fone{}, requiring an improvement over the note-onset baseline and allowing at most 1.0~pp decreases in validation event micro-\Fone{} and frame macro-\Fone{}. This ranking selects joint direct decoding before test evaluation. In the frozen configuration, the original B2 activity predictions are reused directly rather than recomputed by the onset model.

\begin{table}[H]
\centering
\caption{Selected continuation checkpoint epochs and number of epochs run.}
\label{tab:app_epochs}
\small
\setlength{\tabcolsep}{3.5pt}
\begin{tabular}{lrrr}
\toprule
Seed & Context (s) & Best epoch & Epochs run \\
\midrule
42 & 3 & 19 & 24 \\
42 & 18 & 22 & 24 \\
123 & 3 & 1 & 9 \\
123 & 18 & 8 & 16 \\
3407 & 3 & 5 & 13 \\
3407 & 18 & 21 & 24 \\
\bottomrule
\end{tabular}
\end{table}

\section{Detailed Recognition Results}
\label{app:results}
Table~\ref{tab:app_seeds} contains the individual seed results of the fixed-threshold frame comparison (V/T: validation/test; Ma/Mi: macro/micro). Table~\ref{tab:ta02_class_results} gives the class-wise means of the same three configurations on the test split, and Table~\ref{tab:app_precision} gives the precision and recall of the final activity model. The paired results show that the macro-\Fone{} gains from both context extension and overlap aggregation hold for each of the three continuation-stage seeds.

\begin{table}[H]
\centering
\caption{Per-seed activity scores (\%), fixed threshold 0.5.}
\label{tab:app_seeds}
\small
\setlength{\tabcolsep}{3.5pt}
\begin{tabular}{lrrrr}
\toprule
Seed / model & V-Ma & V-Mi & T-Ma & T-Mi \\
\midrule
42 / B0 & 81.98 & 90.48 & 82.35 & 91.74 \\
42 / B1 & 84.84 & 91.85 & 85.89 & 93.08 \\
42 / B2 & 85.05 & 92.37 & 87.88 & 93.55 \\
123 / B0 & 82.23 & 90.50 & 83.98 & 92.02 \\
123 / B1 & 84.64 & 91.44 & 85.55 & 92.84 \\
123 / B2 & 85.14 & 92.19 & 88.39 & 93.47 \\
3407 / B0 & 81.97 & 90.57 & 82.34 & 91.65 \\
3407 / B1 & 85.08 & 92.12 & 85.98 & 93.12 \\
3407 / B2 & 85.48 & 92.57 & 88.06 & 93.34 \\
\bottomrule
\end{tabular}
\end{table}

\begin{table}[H]
\centering
\caption{Class-wise test F1 (\%) and total improvement (pp), three seeds. Merged class: huazhi/guazou/lianmo/liantuo.}
\label{tab:ta02_class_results}
\small
\setlength{\tabcolsep}{3.5pt}
\begin{tabular}{lrrrr}
\toprule
Technique & B0 & B1 & B2 & $\Delta$ (pp) \\
\midrule
chanyin & 84.84 & 85.73 & 86.47 & +1.63 \\
boxian & 95.75 & 96.89 & 96.84 & +1.09 \\
shanghua & 89.22 & 89.67 & 91.19 & +1.97 \\
xiahua & 81.66 & 82.48 & 86.31 & +4.66 \\
Merged class & 93.54 & 91.46 & 93.58 & +0.04 \\
yaozhi & 91.42 & 93.87 & 94.33 & +2.91 \\
dianyin & 43.83 & 60.54 & 68.04 & +24.22 \\
\bottomrule
\end{tabular}
\end{table}

\begin{table}[H]
\centering
\caption{Final activity model (B2): test precision, recall, and F1 (\%), averaged over three seeds at threshold 0.5.}
\label{tab:app_precision}
\small
\setlength{\tabcolsep}{3.5pt}
\begin{tabular}{lrrr}
\toprule
Technique & P & R & F1 \\
\midrule
chanyin & 84.70 & 88.35 & 86.47 \\
boxian & 97.16 & 96.52 & 96.84 \\
shanghua & 95.93 & 86.90 & 91.19 \\
xiahua & 93.18 & 80.39 & 86.31 \\
Merged class & 89.58 & 97.95 & 93.58 \\
yaozhi & 95.95 & 92.78 & 94.33 \\
dianyin & 64.05 & 72.58 & 68.04 \\
\bottomrule
\end{tabular}
\end{table}

\paragraph{Alternative event configurations.}
The development studies consider spectral-flux onset candidates, confidence-based filtering, candidate generation from activity, class-specific onset heatmaps from activity probabilities, and class-specific onset prediction from adapted hidden representations. Table~\ref{tab:app_event_variants} records the corresponding archived test means, including hybrid and routing variants. Candidate generation augments note anchors using the activity sequence; candidate selection filters existing anchors. The activity-heatmap and hidden-heatmap methods learn class-specific onset channels. Hybrid decoding supplements the note-onset baseline with learned candidates; class-wise routing selects a decoding strategy on validation for each technique. The final shared-onset method uses a single class-agnostic onset channel and direct event association. Complementing the candidate interventions in Table~\ref{tab:t04_candidate_diagnostics}, an additional ground-truth diagnostic replaces the entire candidate sequence with annotated reference onsets, yielding 88.27/93.31\% event macro/micro-\Fone{}.

The direct frozen and joint modes are compared at the seed level in Table~\ref{tab:app_event_seeds}; Table~\ref{tab:app_event_class} provides the corresponding class-level event scores. The reference test split contains 1,506 scored events after duplicate removal, including only 11 \textit{yaozhi} starts. Event scores therefore reflect occurrence-level counts rather than active-frame counts, which is important when interpreting the class-wise results.

\newpage
\begin{table}[H]
\centering
\caption{Alternative event configurations in the archived development experiments (test F1, \%; three-seed means).}
\label{tab:app_event_variants}
\small
\setlength{\tabcolsep}{3.5pt}
\begin{tabular}{lrr}
\toprule
Configuration & Macro & Micro \\
\midrule
Spectral-flux candidates & 68.72 & 88.30 \\
Union with NMS & 76.49 & 89.89 \\
Confidence filtering & 78.02 & 91.38 \\
Joint filtering & 76.97 & 90.73 \\
Yaozhi recovery & 79.33 & 91.11 \\
Candidate selection & 74.19 & 92.12 \\
Candidate generation & 79.92 & 91.16 \\
Selection + generation & 77.17 & 92.10 \\
Activity heatmap, direct & 68.77 & 56.20 \\
Activity heatmap, hybrid & 79.81 & 91.18 \\
Hidden heatmap, direct & 76.65 & 92.64 \\
Hidden heatmap, hybrid & 80.23 & 91.23 \\
Hidden heatmap, routed & 79.76 & 93.68 \\
Frozen IPT, hybrid & 80.07 & 91.27 \\
Joint IPT, hybrid & 79.66 & 91.07 \\
\bottomrule
\end{tabular}
\end{table}

\begin{table}[H]
\centering
\caption{Per-seed test event scores (macro/micro, \%).}
\label{tab:app_event_seeds}
\small
\setlength{\tabcolsep}{3.5pt}
\begin{tabular}{lrrr}
\toprule
Seed & Note onset & Decoupled & Joint \\
\midrule
42 & 78.99 / 91.18 & 80.79 / 92.66 & 80.35 / 92.63 \\
123 & 78.97 / 91.08 & 81.17 / 92.68 & 81.20 / 92.88 \\
3407 & 79.57 / 91.27 & 81.23 / 92.87 & 81.60 / 92.76 \\
\bottomrule
\end{tabular}
\end{table}

\begin{table}[H]
\centering
\caption{Class-wise test event F1 (\%), averaged over three seeds.}
\label{tab:app_event_class}
\small
\setlength{\tabcolsep}{3.5pt}
\begin{tabular}{lrrr}
\toprule
Technique & Note onset & Decoupled & Joint \\
\midrule
chanyin & 85.51 & 86.02 & 86.38 \\
boxian & 94.21 & 95.94 & 96.18 \\
shanghua & 90.87 & 94.04 & 93.18 \\
xiahua & 89.66 & 89.66 & 89.66 \\
Merged class & 88.61 & 87.66 & 86.02 \\
yaozhi & 37.12 & 45.88 & 47.81 \\
dianyin & 68.25 & 68.25 & 68.14 \\
\bottomrule
\end{tabular}
\end{table}

\end{document}